\documentclass[12pt]{revtex4}
\pdfoutput=1
\usepackage[final]{pdfpages}
\usepackage{amsmath}
\usepackage{fancyhdr}
\usepackage{graphicx}
\usepackage{ifthen}
\usepackage[final]{pdfpages}
\usepackage{titlesec}
\usepackage{array}
\usepackage{multirow}
\titleformat{\section}{\large\bfseries}{\thesection}{1em}{}

\numberwithin{equation}{section}

\begin{document}



# Calculations of Nonlinear Wave-Packet Interferometry Signals in the Pump-Probe Limit as Tests for Vibrational Control over Electronic Excitation Transfer

Jason D. Biggs and Jeffrey A. Cina[*]

Department of Chemistry and Oregon Center for Optics,

University of Oregon, Eugene, Oregon, 97403

**Abstract.** The preceding paper [J. D. Biggs and J. A. Cina, xxx; referred to here as Paper 1], describes a strategy for externally influencing the course of short-time electronic excitation transfer (EET) in molecular dimers and observing the process by nonlinear wave-packet interferometry (nl-WPI). External influence can, for example, be exerted by inducing coherent intramolecular vibration in one of the chromophores prior to short-pulse electronic excitation of the other. Within a sample of isotropically oriented dimers having a specified internal geometry, a vibrational mode internal to the acceptor chromophore can be preferentially driven by electronically nonresonant impulsive stimulated Raman (or resonant infrared) excitation with a short polarized "control" pulse. A subsequent electronically resonant polarized pump then preferentially excites the donor, and EET ensues. Paper 1 investigates control-pulse-influenced nl-WPI as a tool for the spectroscopic evaluation of the effect of coherent molecular vibration on excitation transfer, presenting general expressions for the nl-WPI difference signal from a dimer following the action of a control pulse of arbitrary polarization and shape.

---

[*] Electronic mail: cina@uoregon.edu

Electronic excitation is to be effected and its inter-chromophore transfer monitored by resonant pump and probe "pulses," respectively, each consisting of an optical-phase-controlled ultrashort pulse-pair having arbitrary polarization, duration, center frequency, and other characteristics. Here we test both the control strategy and its spectroscopic investigation—with some sacrifice of amplitude-level detail—by calculating the pump-probe difference signal. That signal is the limiting case of the control-influenced nl-WPI signal in which the two pulses in the pump pulse-pair coincide, as do the two pulses in the probe pulse-pair. We present calculated pump-probe difference signals for (1) a model excitation-transfer complex in which two equal-energy monomers each support one moderately Franck-Condon active intramolecular vibration; (2) a simplified model of the covalent dimer dithia-anthracenophane, representing its EET dynamics following selective impulsive excitation of the weakly Franck-Condon active $\nu_{12}$ anthracene vibration at 385 cm$^{-1}$; and (3) a model complex featuring moderate electronic-vibrational coupling in which the site energy of the acceptor chromophore is lower than that of the donor.

## 1. BACKGROUND & SET-UP

Recent theoretical studies of electronic excitation transfer (EET) have explored the influence of coupling between the purely electronic degrees of freedom and "environmental" nuclear motion (both intra- and intermolecular). Making use of experimentally guided choices for bath-mode correlation functions, Aspuru-Guzik and co-workers[1,2] performed computations tracking useful measures for the influence of different environment-induced processes on energy transfer. Plenio *et al.*[3,4] subsequently analyzed the fundamental effects of electronic dephasing, relaxation, and trapping within

the context of models relying on a phenomenological Markovian description of the environmental interactions. Both groups have made important headway in elucidating the observed interplay among coherent electronic dynamics, medium-induced decoherence and relaxation, and trapping processes in complex multi-chromophore systems such as the Fenna-Matthews-Olson complex,[5, 6] bacterial photosynthetic reaction centers,[7] and conjugated polymers.[8] In the context of these studies, the electronic degrees of freedom are regarded as the "system," while the "bath" comprises both inter- and intramolecular vibrational degrees of freedom.

Other studies have addressed certain manifestly non-Markovian effects that can occur in the presence of fairly strong coupling between electronic excitations and intramolecular or local vibrations. Jang, Jung, and Silbey[9] formulated a time-dependent generalization of Förster theory accounting for the effects on energy transfer of coherent vibrational motion in a Franck-Condon active mode. Cina and Fleming[10] performed model-system calculations seeking to rationalize the observation of vibrational coherence transfer in time-resolved polarized fluorescence up-conversion measurements on LH-1.[11, 12]

In the preceding paper[13] (referred to as Paper 1), Biggs and Cina describe a possible means of exerting external influence over the time-course of initial electronic excitation transfer by inducing a coherent intramolecular vibration in the acceptor chromophore prior to short-pulse excitation of the donor. Their strategy builds on the work of Jang, Jung, and Silbey[9] and Cina and Fleming,[10] and seeks to control the instantaneous "rate" of excitation transfer by initiating vibrational motion in a Franck-

Condon active acceptor mode in order to render time-dependent the acceptor absorption frequency relative to the emission frequency of the donor.

Motivated by current ultrafast experimental capabilities,[5,8,14,15,16,17,18,19,20,21] this recent theoretical work generalizes Förster's foundational insights on excitation transfer,[22] and builds on other detailed studies of the dynamics of few- or multi-level quantum mechanical systems in condensed media.[23,24,25,26,27,28]

Paper 1 presents general expressions for the nl-WPI difference signal from an EET dimer following the action of a control pulse of arbitrary polarization and shape. The signal expressions given there apply to a sample of isotropically distributed dimers, having a specified internal geometry, in which the acceptor vibration can be preferentially driven by impulsive stimulated Raman excitation with a short, polarized "control" pulse.[29,30,31] An electronically resonant, polarized "pump" comprising two optical-phase-related pulses then preferentially excites the donor, and the action of a polarized phase-controlled "probe" pulse-pair enables the detection of a wave-packet interferometry signal. Such control-pulse-influenced nl-WPI measurements will provide highly detailed nuclear-amplitude-sensitive records of the effect of the externally induced nuclear motion on the time-course of EET. Here, we perform numerical tests of this approach to the control and observation of short-time EET by applying the previous general treatment to some simplified model systems, and by collapsing the pump and probe pulse-pairs in the WPI sequence each to a single pump or probe pulse.

Paper 1 treats a molecular dimer with a field-free Hamiltonian

$$H = |0\rangle H_0 \langle 0| + |1\rangle H_1 \langle 1| + |1'\rangle H_{1'} \langle 1'| + |2\rangle H_2 \langle 2| + J\{|1'\rangle\langle 1| + |1\rangle\langle 1'|\} \ . \quad (1)$$

$H_n$ ( $n = 0,\ 1,\ 1',\ 2$ ) is the nuclear Hamiltonian in the uncoupled electronic state $|n\rangle$,

where

$$|0\rangle = |gg\rangle \qquad \text{(both monomers unexcited)} \qquad (2)$$

$$|1\rangle = |eg\rangle \qquad \text{("donor" excited)} \qquad (3)$$

$$|1'\rangle = |ge\rangle \qquad \text{("acceptor" excited)} \qquad (4)$$

$$|2\rangle = |ee\rangle \qquad \text{(both monomers excited)}\ , \qquad (5)$$

and $J$ is a parameter determining the strength of the energy-transfer coupling (taken here to be constant). The designation of one molecule as the donor and the other as the acceptor is arbitrary in the case of a homo-dimer of monomers with equal site energy. The signal calculations shown below for such systems treat both chromophores identically; their excitation and de-excitation probabilities depend upon the direction cosines between their individual transition dipole moments and the various field polarizations. In calculations of time-resolved signals from hetero-dimers, the donor (acceptor) will be the monomer having the higher (lower) site energy.

The Hamiltonian (1) is general in form, allowing any number of intramolecular vibrational and bath degrees of freedom—including "mode-sharing" between the monomers—but we specialize here to some simple cases allowing illustrative tests of the basic control strategy and its spectroscopic verification. We ascribe to each chromophore a single internal harmonic vibration whose equilibrium coordinate value is displaced in the monomer's $e$-state, but whose frequency is the same for both monomers and remains unchanged upon electronic excitation. The nuclear Hamiltonians can then be written as

$$H_0 = \frac{p_a^2 + p_b^2}{2m} + \frac{m\omega^2}{2}\left[ q_a^2 + q_b^2 \right] \qquad (6)$$

$$H_1 = \frac{p_a^2 + p_b^2}{2m} + \frac{m\omega^2}{2}\left[(q_a - d)^2 + q_b^2\right] + \varepsilon_1 \tag{7}$$

$$H_{1'} = \frac{p_a^2 + p_b^2}{2m} + \frac{m\omega^2}{2}\left[q_b^2 + (q_b - d)^2\right] + \varepsilon_{1'} \tag{8}$$

$$H_2 = \frac{p_a^2 + p_b^2}{2m} + \frac{m\omega^2}{2}\left[(q_a - d)^2 + (q_b - d)^2\right] + \varepsilon_2 \quad . \tag{9}$$

$\omega$ is the common vibrational frequency of both intramolecular modes; $\varepsilon_1$ and $\varepsilon_{1'}$ are the donor and acceptor site energies, respectively; $\varepsilon_2$ is the energy of the two-exciton state (perhaps roughly equal to $\varepsilon_1 + \varepsilon_{1'}$); and $d$ is the displacement of the $e$-state nuclear potential curves (of importance in the calculations performed below is the dimensionless quantity $\delta = d\sqrt{m\omega/2}$, the coordinate displacement $d$ divided by $2\Delta q_{rms} = \sqrt{2/m\omega}$; $\hbar \equiv 1$ throughout). $q_{a(b)}$ and $p_{a(b)}$ are the position and momentum operators for a vibrational mode localized on the donor (acceptor) chromophore respectively. Extending the Hamiltonian (1) to multiple chromophores recovers the Frenkel exciton Hamiltonian widely used to study energy migration in molecular aggregates.[32,33,34]

Calculations are performed using the eigenbasis of the uncoupled ($J = 0$) Hamiltonian; $|n\rangle |(\nu_a \nu_b)_n\rangle$ denotes the state in which the electronic degrees of freedom are in state $n$, while the vibrations are in an eigenstate of $H_n$ with $\nu_a$ and $\nu_b$ quanta in the donor and acceptor, respectively. The basis is truncated beyond states having more than a certain total number of vibrational quanta (between 10 and 24 for the parameter values used here). Free evolution under the EET-coupled Hamiltonian (1) is treated by numerical matrix diagonalization to obtain a representation of

$$[t] \equiv e^{-iHt} = \sum_j |\Psi_j\rangle e^{-iE_j t} \langle \Psi_j| \tag{10}$$

in our finite basis, where $\left| \Psi_j \right\rangle$ is an eigenket of $H$ with energy $E_j$.

We wish to predict the signals from ultrafast pump-probe difference measurements on the system described by Eq. (1). The measured observable is the contribution to the population of the one-exciton manifold (states 1 and 1′) that is simultaneously linear in the intensities (electric field squared) of a nonresonant control pulse (denoted by $P$), an electronically resonant pump pulse ($A$), and a resonant probe pulse ($C$); see Paper 1 for more details. The electric field of the pulse $I$ has the form specified by Eq. (5) of Paper 1, with a Gaussian envelope function

$$f_I(t) = \exp\left\{ -(t - t_I)^2 / 2\sigma_I^2 \right\} \tag{11}$$

and a phase function

$$\Phi_I(t) = \Omega_I \cdot (t - t_I) \quad . \tag{12}$$

The first- and second-order interactions of the system with each nonzero-duration laser pulse are described by pulse-propagators (see Eqs. (12) and (17) of Paper 1, and further below). These are matrices—whose elements need be calculated only once—which describe a pulse's effect on the state of the system irrespective of the variable arrival time $t_I$. Calculation of the pulse-propagators is simplified by assuming that the pulse duration $\sigma_I$ is short compared to the characteristic timescale $2\pi/J$ for excitation transfer, so that EET can be ignored in accounting for the action of a pulse. Using the pulse shape (11) and phase function (12), we find, for example,

$$\left\langle (\nu_a, \nu_b)_1 \left| p_I^{(10)}(\infty) \right| (\overline{\nu}_a, \overline{\nu}_b)_0 \right\rangle = i\sqrt{\frac{\pi}{2}} E_I m \sigma_I \left\langle (\nu_a)_e \left| (\overline{\nu}_a)_g \right\rangle \delta_{\nu_b, \overline{\nu}_b} \right.$$
$$\times \exp\left\{ -\frac{\sigma_I^2}{2} \left( \Omega_I + \omega(\overline{\nu}_a - \nu_a) - \varepsilon_1 \right)^2 \right\} , \tag{13}$$

in the notation of Eq. (14) in Paper 1; $\delta_{\nu_b \bar{\nu}_b}$ is a Kronecker delta.[35] Other matrix elements of the first-order pulse propagators can be similarly obtained. The Franck-Condon overlap appearing on the RHS of Eq. (13) can be calculated straightforwardly using displacement-operator methods.[36]

The second-order action of a pulse (such as the control pulse) includes contributions from matrix elements of the type

$$
\left\langle (\nu_a, \nu_b)_0 \left| p_I^{(01)}(\infty; t_2) p_I^{(10)}(t_2; t_1) \right| (\bar{\nu}_a, \bar{\nu}_b)_0 \right\rangle = \left( \frac{iE_I m}{2} \right)^2 \int_{-\infty}^{\infty} dt_2 \int_{-\infty}^{t_2} dt_1 \, f_I(t_2) f_I(t_1) e^{i\Phi_I(t_2) - i\Phi_I(t_1)}
$$
$$
\times \left\langle (\nu_a, \nu_b)_0 \left| e^{-iH_0(t_I - t_2)} e^{-iH_1(t_2 - t_I)} e^{-iH_1(t_I - t_1)} e^{-iH_0(t_1 - t_I)} \right| (\bar{\nu}_a, \bar{\nu}_b)_0 \right\rangle . \tag{14}
$$

Using the eigenenergies of ground and one-exciton vibronic states, along with the Gaussian pulse envelope (11) and the phase function (12), and making a sequence of integration-variable changes, Eq. (14) can be evaluated analytically to give

$$
\left\langle (\nu_a, \nu_b)_0 \left| p_I^{(01)}(\infty; t_2) p_I^{(10)}(t_2; t_1) \right| (\bar{\nu}_a, \bar{\nu}_b)_0 \right\rangle
$$
$$
= -\frac{\pi}{4} \sigma_I^2 E_I^2 m^2 \delta_{\nu_b \bar{\nu}_b} \sum_{\bar{\nu}_a} \left\langle (\nu_a)_g \left| (\bar{\bar{\nu}}_a)_e \right\rangle \left\langle (\bar{\bar{\nu}}_a)_e \left| (\bar{\nu}_a)_g \right\rangle \right.
$$
$$
\times \exp\left\{ -\sigma_I^2 \Delta^2 - \frac{\sigma_I^2 \omega^2}{4} (\nu_a - \bar{\nu}_a)^2 \right\} \left[ 1 + \text{erf}\left( -i\sigma_I \Delta \right) \right] , \tag{15}
$$

where

$$
\Delta \equiv \Omega_I + \omega \left( \frac{\nu_a + \bar{\nu}_a}{2} - \bar{\bar{\nu}}_a \right) - \varepsilon_1 . \tag{16}
$$

Appendix B (online)[37] details the derivation of Eq. (15) and other elements of the second-order pulse propagators. Possession of a compact expression for the second-order action of a pulse propagator—whose identification is a key practical result of this paper—greatly facilitates the calculation of short-pulse nonlinear optical signals by eliminating the need for repeated numerical evaluation of nested time-integrals like those in Eq. (14).

8

Similar expressions can also be derived for linearly chirped Gaussian pulses, whose phase functions are quadratic in time.[38] Although results of this kind might appear to be limited to displaced harmonic systems, they can also be applied to arbitrary anharmonic potentials by replacing the latter with harmonic approximations whose value, slope, and curvature match those of the actual potential at the instantaneous location of the wave packet on which the propagators act.[39]

## 2. CALCULATIONS

### A. Model system with moderate electronic-vibrational coupling

We first consider a model system with moderate vibrational displacements, $\delta = \sqrt{2.5}$, in the site-excited states; a value $J = 0.2\,\omega$ for the energy-transfer parameter; and perpendicular donor and acceptor electronic transition-dipole moments. The monomers are assigned equal site energies, $\varepsilon_1 = \varepsilon_{1'} \equiv \varepsilon$, and $\varepsilon_2 = \varepsilon_1 + \varepsilon_{1'} = 2\varepsilon$.[40] A numerical search over pulse duration and center frequency, subject to a constraint of less than 5% electronic excitation probability, found that among transform-limited Gaussian control pulses the combination $\sigma_P = 0.140\,(2\pi/\omega)$ and $\Omega_P = \varepsilon - 1.91\,\omega$ generates the largest amplitude of nuclear motion, $0.338\,d = 1.068\,\Delta q_{rms}$, in the electronic ground state. All subsequent calculations ignore the small contribution due to control-induced electronic excitation. In order that the control pulse selectively excite vibrational motion in the "acceptor" chromophore, while the pump and probe pulses excite and probe the "donor," we choose a VHH polarization combination in which the control pulse is perpendicular to the mutually parallel pump and probe. The pump pulse, with duration $\sigma_A = 0.1\,(2\pi/\omega)$ and center frequency $\Omega_A = \varepsilon + \delta^2\,\omega$, is locally resonant with the $1 \leftarrow 0$

transition along the line $q_a = 0$ and with $1' \leftarrow 0$ along $q_b = 0$. The probe pulse, with $\sigma_C = 0.25(2\pi/\omega)$ and $\Omega_C = \varepsilon - 3\delta^2 \omega$, is resonant with $1 \rightarrow 0$ along the line $q_a = 2d$ and with $1' \rightarrow 0$ along the line $q_b = 2d$. These lines are the loci of vibrational outer turning points in the 1 and $1'$ states, respectively.

Figure 1 shows the fractional population of the one-exciton manifold residing in the donor state (state-1) of an oriented complex (one site transition-dipole vertical, the other horizontal) following $A$-pulse excitation to that state from the electronic ground state without and with prior impulsive-Raman excitation of coherent vibrational motion in the acceptor chromophore at time $t_P = t_A - 0.56(2\pi/\omega)$ (chosen so that the ground-state wave packet is at its inner turning point, with $\langle q_b \rangle = -0.338d$, when the $A$ pulse arrives). Despite the fact that the acceptor-mode displacement achieved by the impulsive Raman process is smaller than the ideal value $\langle q_b \rangle = -d$ envisaged in Fig. 1 of Paper 1, the control pulse evidently slows excitation transfer significantly during the first seven vibrational periods.

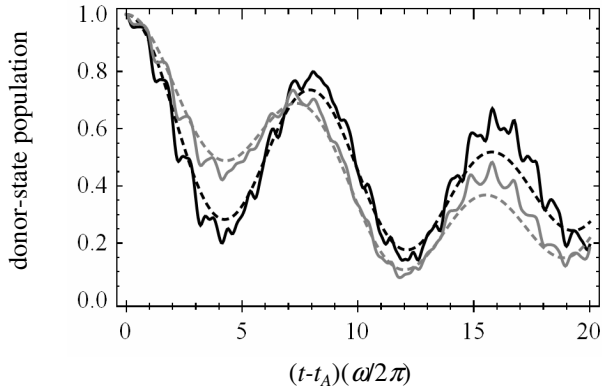

FIG. 1. Relative population of donor state versus time (in vibrational periods) after electronic excitation from vibrational ground state (solid black curve) and from impulsively displaced vibrational wave packet (solid gray curve). Corresponding dashed curves are semi-analytical predictions of weak EET-coupling

theory derived in Appendix A, which depend only on vibronic state populations and miss effects due to vibrational coherence.

Figure 2 compares the calculated pump-probe and pump-probe difference signals for this system in an isotropic sample. In the absence of influence by control-pulse vibrational stimulation on the time-course of excitation transfer, these two signals would be proportional to each other at all times. The lack of simple proportionality between the two curves in Fig. 2 therefore demonstrates coherent control of EET. Both the survival probabilities and the pump-probe signals are influenced by vibrational coherences. The latter spectroscopic signals, though, are additionally complicated because the signal selects for nuclear probability density in a specified probe-window region of vibrational configuration space. Thus the pump-probe signals provide detailed views of the nuclear *and* electronic dynamics in this EET system, as they reflect the instantaneous spatial probability density in the probe-window region, rather than just the total population of a specific electronic state.[41,42] In this model system with moderately strong electronic-vibrational coupling, the envelope of the probe-probe signal and the pump-probe difference signal nevertheless roughly resemble the corresponding population traces of Fig. 1.[43] In particular, the envelope of the pump-probe difference signal decreases less precipitously during the first $\sim 5(2\pi / \omega)$ than the envelope of the pump-probe signal without a control pulse.

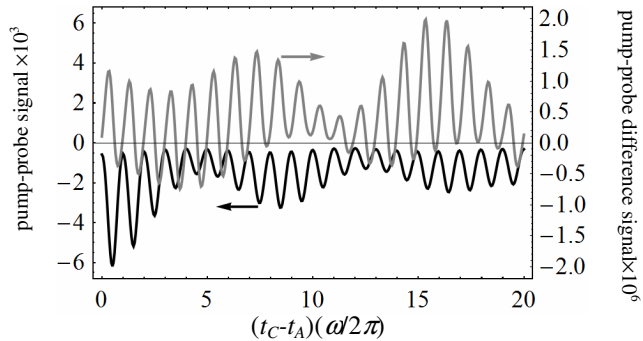

FIG. 2. Pump-probe signal (black curve, left ordinate in units of $E_A^2 E_C^2 m^4 / 2^4$ ) and pump-probe

difference signal (gray curve, right ordinate in units of $E_P^2 E_A^2 E_C^2 m^6 / 2^6$ ) as a function of delay time

between pump and probe pulses. Signals are calculated for an isotropic sample of the model energy-transfer

complex with equal-energy monomers having perpendicular transition moments.

Because the probe-pulse center frequency is strongly red-shifted from the pump

( $\Omega_C - \Omega_A = -4\delta^2 \omega$ ), both signals in Fig. 2 are dominated by their stimulated-emission

contributions. Those contributions (not shown separately)—specified for the pump-probe

difference by Eq. (42) and Appendix C (online) of Paper 1—are indistinguishable on the

scale of Fig. 2 from the total signals including also excited-state absorption and ground-

state bleaching.[44]

The presence of six direction cosines (see Appendix A of Paper 1) suggests that

the VHH-polarized pump-probe difference signal from an isotropic sample should be

dominated by those terms which would be nonzero if the EET complex were oriented

with one transition moment along the vertical axis and the other horizontal. In Figure 3

are plotted those portions of the stimulated-emission signals arising from the wave-packet

overlaps that would be generated in such ideally oriented complexes.[45] The handful of

overlaps contributing to the signals in Fig. 3, which is plotted on the same scale as Fig. 2,

evidently account a major portion of the pump-probe and pump-probe difference signals

from the isotropic sample.

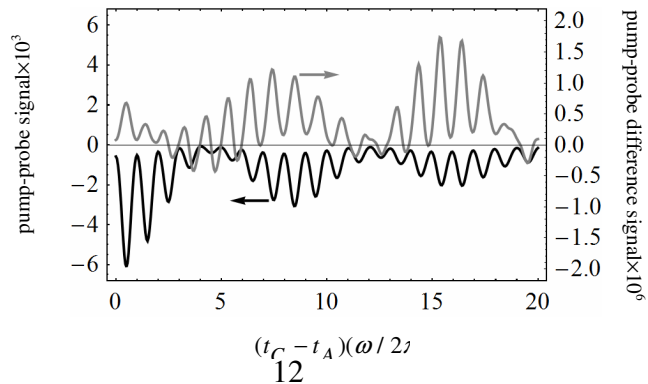

FIG. 3. Stimulated-emission contributions to the pump-probe and pump-probe difference signals shown in Fig. 2 from those wave-packet overlaps that would take nonzero values if one monomer's transition moment were oriented vertically (parallel to the control-pulse polarization) and the other's were oriented horizontally (parallel to the pump and probe polarization).

Although the main focus here is on the energy-transfer dynamics of putatively isolated systems, some indication of the effects of medium-induced dephasing on the calculated signals can be gained by introducing electronic inhomogeneous broadening.[10] To this end, we average the signals over a collection of isotropic dimer systems in which monomer site energies, $\varepsilon_1$ and $\varepsilon_{1'}$, are independently distributed about a common mean value $\varepsilon$. The less rapid decrease in VHH pump-probe difference signal compared to the HH pump-probe signal seen in Fig. 2 remains evident in the inhomogeneously broadened signals shown in the upper panel of Figure 4.[46] The corresponding perpendicular-probe cases, VHV and HV are shown there as well, and the resulting signal anisotropies— (VHH-VHV)/(VHH+2VHV) for pump-probe difference and (HH-HV)/(HH+2HV) for pump-probe—are plotted in the lower panel of Fig. 4. There is a notably lower initial value in the control-influenced pump-probe difference anisotropy, but it does not appear to decay less rapidly than the pump-probe anisotropy for this system. Appendix C (online)[47] addresses the interesting dependence of the initial pump-probe difference anisoptropy on the duration of pulses $A$ and $C$. But the findings presented there must be considered approximate because the theory of Paper 1 (see its Theory section) and the calculations of this paper neglect the effects of temporal $A$-$C$ overlap. Such pump-probe overlap effects are unlikely to be entirely negligible when $t_C - t_A = 0$.

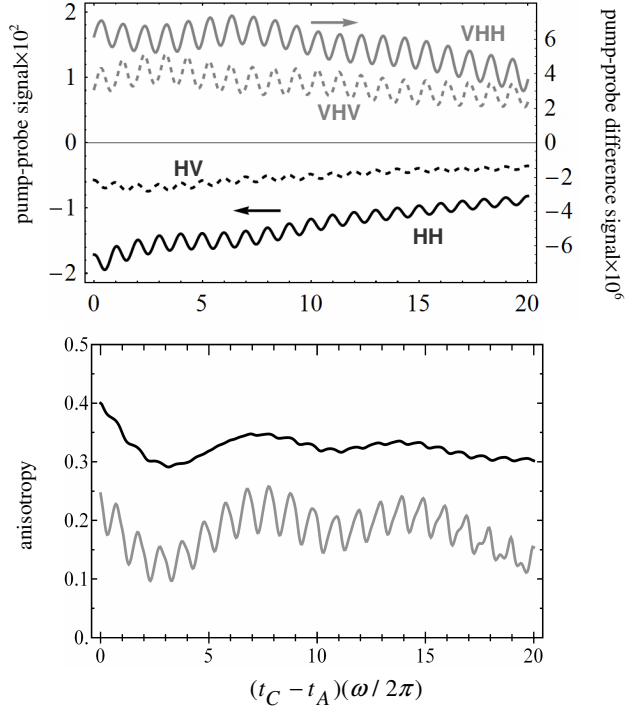

FIG. 4. Upper panel: Average signals from a collection of 1000 model-system dimers ( $\delta = \sqrt{2.5}$ , $J = 0.2\omega$ ), whose site energies are chosen from independent normal distributions ( $\bar{\varepsilon}_1 = \bar{\varepsilon}_{1'} \equiv \varepsilon$ ) with FWHM equal to the vibrational frequency. Pulse parameters are the same as in Fig. 2, except the temporal widths of the pump and probe pulses are $\sigma_A = 0.25 \, (2\pi/\omega)$ and $\sigma_C = 0.5 \, (2\pi/\omega)$ , respectively. Pump-probe signals are shown for HH and HV polarizations; pump-probe difference signals for VHH and VHV polarizations. Lower panel: Signal anisotropies for the inhomogeneously broadened model system, in black for pump-probe and gray for pump-probe difference.

## B. Dithia-anthracenophane

Next we examine excitation-transfer dynamics and pump-probe signals from an EET complex with parameter values appropriate to dithia-anthracenophane (DTA), a covalent anthracene dimer compound that has been studied by Yamazaki and co-workers using time-resolved pump-probe anisotropy measurements.[48,49] Two vibrational modes of the anthracene monomer, mode-12 at $\omega_{12}/2\pi c = 385 \, \mathrm{cm}^{-1}$ and mode-6 at $\omega_6/2\pi c \cong 1400 \, \mathrm{cm}^{-1}$ , dominate the vibronic progression of its $S_1 \leftarrow S_0$ fluorescence-

14

excitation spectrum within $1500\,\text{cm}^{-1}$ of its origin at $27695\,\text{cm}^{-1}$.[50] From the jet-spectroscopy data of Lambert *et al.* we abstract a dimensionless displacement $\delta_{12} = \sqrt{0.31}$ for the lower-frequency vibration. We use an effective energy-transfer coupling constant $J = (33.6\,\text{cm}^{-1}/2)\exp(\delta_{12}^2) = 0.0595\ \omega_{12} = 22.9\ \text{cm}^{-1}$ obtained by dividing half the population-oscillation frequency reported in Ref. 49 by the estimated Franck-Condon overlap between two mode-12 wave packets separated by their relative displacement $(\sqrt{2}d_{12})$ in the two site-excited potential wells.[51] The two anthracene-monomer transition dipole moments in DTA are known to be approximately, but not exactly, orthogonal;[48] for simplicity, their relative angle is set to $\alpha = \pi/2$ in our calculations.

In the calculations shown here, we have taken chosen pulse widths and center frequencies so that only superpositions of mode-12 vibronic states are prepared upon electronic excitation and only mode-12 is susceptible to impulsive-Raman excitation in the electronic ground state. In particular, mode-6 is quiescent under the action of both the control pulse ($\sigma_P = 0.225(2\pi/\omega_{12}) = 19.5\,\text{fs}$ ; $\Omega_P = \varepsilon - 1.53\omega_{12}$) and the vertically resonant pump ($\sigma_A = 0.1(2\pi/\omega_{12}) = 8.66\,\text{fs}$ ; $\Omega_A = \varepsilon + \delta_{12}^2 \omega_{12}$). We therefore report calculations on a simplified model, referred to as DTA-12, in which only this single Franck-Condon active mode is treated explicitly. The probe pulse is longer than the pump and resonant at the outer turning line of nuclear motion in either site-excited state of DTA-12, with $\sigma_C = 0.5(2\pi/\omega_{12}) = 43.3\,\text{fs}$ and $\Omega_C = \varepsilon - 3\delta_{12}^2 \omega_{12}$. A VHH polarization scheme is adopted (as explained in subsection 2A). The control pulse generates mode-12 nuclear motion in the electronic ground state with amplitude $0.54d_{12} = 0.60\Delta q_{12}$. The

delay between the control and pump pulses, $0.56(2\pi / \omega_{12}) = 48.5\,\text{fs}$, is chosen so that this control-induced wave packet reaches its inner turning point, $\langle q_b \rangle = -0.54 d_{12}$, at the arrival time of the pump pulse.

Calculated pump-probe and pump-probe difference signals from an isotropic DTA-12 sample are shown in the upper panel of Figure 5. Due to the small $\delta_{12}$ value, both excited-state absorption and ground-state bleach, along with stimulated emission, contribute significantly to the calculated signals. One-exciton emission signals very nearly equivalent to the stimulated emission components[52] of the pump-probe and pump-probe difference signals (Fig. 5, lower panel) could however be isolated in fluorescence up-conversion measurements.[53]

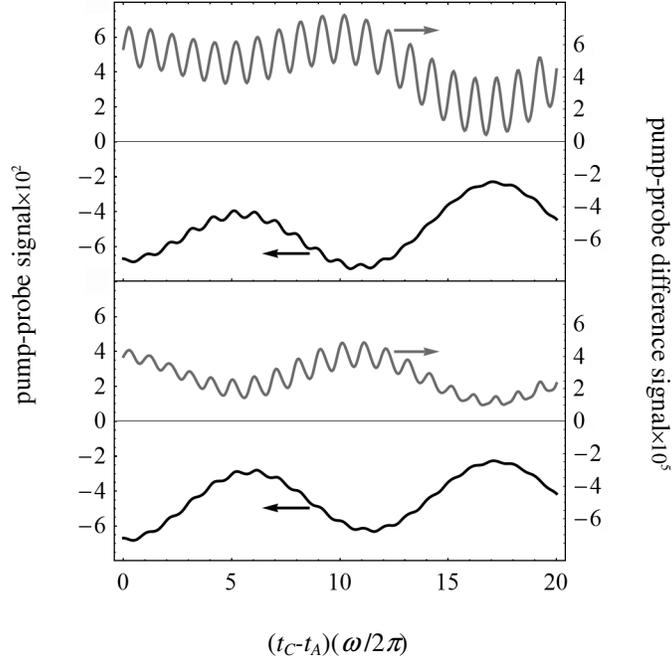

FIG. 5. Pump-probe signal (black curve, left ordinate) and pump-probe difference signal (gray curve, right ordinate) as function of the delay between pump and probe pulses. Upper panel shows pump-probe and pump-probe difference signals from simplified model of dithia-anthracenophane treating only the dynamics of the low-frequency mode-12. Lower panel plots stimulated-emission contributions from DTA-12.

Despite their general similarity in form, the pump-probe difference and corresponding pump-probe signals in Fig. 5 are not constant multiples of each other, so the nuclear motion induced by the control pulse evidently affects to some degree the wave-packet surface-crossing dynamics in the one-exciton manifold. We turn to the population dynamics from an oriented DTA-12 complex to gain further insight. Figure 6 plots the donor-state survival probability following the action of the control and pump pulses on a dimer with one site transition-dipole aligned with the control-pulse polarization (vertical) and the other aligned with the pump (horizontal). The control pulse induces small-amplitude motion in the acceptor— $0.54 d_{12} = 0.54\delta\sqrt{2/m\omega_{12}}$ $= 0.60/\sqrt{2m\omega_{12}}$, less than the rms width of the wave packet—so the population dynamics

in the presence of the control pulse differs only slightly from that in its absence. Moreover, the small-amplitude vibration *accelerates* short-time electronic excitation transfer in DTA-12 rather than forestalling it, by increasing the amplitude of the donor-to-acceptor population oscillation. As explained further in Appendix A, this behavior results from the fact that the vibrational displacement in mode-12 is not large enough to remove the donor-state wave packet from the region of intersection between the donor-excited and acceptor-excited potential energy surfaces. The quasi-classical argument suggesting that acceptor-mode displacement should impede short-time EET (see Fig. 1 of Paper 1 and the accompanying explanation) does not apply in this situation.

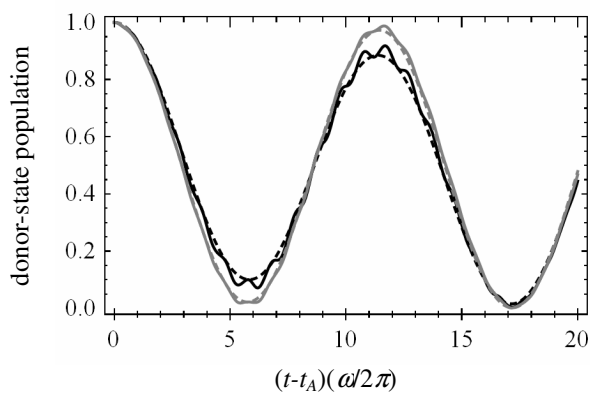

FIG. 6. Donor-excited-state population dynamics for DTA-12 without (solid black curve) and with (solid gray curve) prior impulsive excitation of mode-12 vibration in acceptor chromophore. Transition dipole moment of acceptor chromophore is aligned with vertical control-pulse polarization. Transition dipole moment of donor is aligned with horizontal pump-pulse polarization. Dashed curves show corresponding predictions of weak EET-coupling approximation (see Appendix A).

In an effort to model the effects of electronic inhomogeneous broadening in DTA-12, we can introduce variable monomer site energies (as described in subsection 2A). Figure 7 shows the pump-probe and pump-probe difference anisotropies obtained from the inhomogeneously broadened stimulated-emission components. The calculated signals

from which these anisotropies were obtained were averaged over 1200 DTA-12 dimers whose site energies were chosen from independent Gaussian distributions with 100-cm$^{-1}$ FWHM. Pulse parameters are the same as in Fig. 5, except $\sigma_A = 0.25\,(2\pi/\omega_{12}) = 21.6\,\text{fs}$. The slight control-induced acceleration of EET seen in the lower panel of Fig. 5 is discernible also in the shorter-period anisotropy oscillation from the pump-probe difference than the pump-probe signal in Fig. 7.

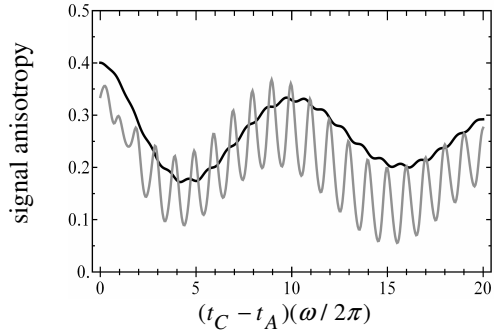

FIG. 7 Anisotropies calculated from inhomogeneously broadened stimulated emission component of the pump-probe signal (black) and pump-probe difference signal (gray).

## C. Downhill energy transfer

The calculations of subsections 2A and 2B focus on homodimers, in which the site excitation energies of the participating monomers are the same. We next investigate the vibrational influence on EET in a heterodimer whose excitation donor has a higher site energy than the acceptor. In order to facilitate comparison with the calculations of subsection 2A, we use the same values of all system parameters except $\varepsilon_{1'} = \varepsilon_1 - 2\omega\delta^2$. This difference in site energies is chosen so that the state-1′ potential energy surface passes through the minimum of the state-1 potential located at $(q_a, q_b) = (d, 0)$. The diagonal intersection line between donor-excited and acceptor-excited potentials also passes through the point $(q_a, q_b) = (0, -d)$ corresponding to "ideal" acceptor-mode

19

displacement. In this "downhill" model, it is expected that electronic excitation of the donor chromophore at an instant when the coherently vibrating acceptor mode is at its inner turning point should lead to accelerated short-time energy transfer.

The control pulse used to effect this coherent impulsive excitation has the same vertical polarization and pulse duration ($\sigma_P = 0.140\,(2\pi/\omega)$) as in subsection 2A. Its center frequency is down-shifted by $2\omega\delta^2$ to $\Omega_P = \varepsilon_{1'} - 1.91\omega = \varepsilon_1 - 6.91\omega$ in keeping with the lower acceptor site energy, and the acceptor-mode wave packet is therefore identical to that considered previously. The horizontally polarized pump pulse has the same center frequency ($\Omega_A = \varepsilon_1 + \delta^2\,\omega$), duration ($\sigma_A = 0.1\,(2\pi/\omega)$), and arrival time relative to the control pulse as before; it is locally resonant with $1 \leftarrow 0$ along $q_a = 0$ and with $1' \leftarrow 0$ along $q_b = -d$, which is largely unpopulated by nuclear probability amplitude belonging to either the vibrational ground state or the impulsively excited vibrational wave packet in the electronic ground state (given the $0.338\,d$ range of control-induced vibrational motion).

Calculations of the donor-state survival probability for a downhill complex with a vertically (horizontally) aligned acceptor (donor) transition moment are shown in Figure 8. The results bear out the expectation that the initial acceptor-mode displacement should accelerate short-time excitation transfer and accentuate oscillatory forward-and-backward EET (of period $\sim 6(2\pi/\omega)$).

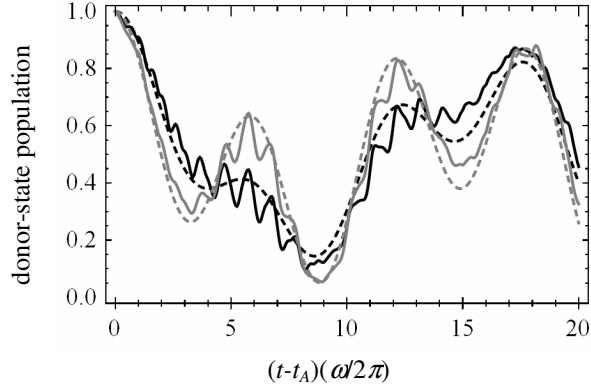

FIG. 8. Donor-state survival probabilities for oriented downhill EET model without (solid black curves) and with (solid gray curves) prior impulsive excitation of acceptor-mode vibration. Coherent vibrational excitation of acceptor is seen to accelerate short-time EET in this system. Dashed curves give the corresponding predictions under weak EET-coupling approximation.

For the downhill system, we investigate pump-probe and pump-probe difference signals using probe pulses of two different kinds. In order selectively to detect primarily the nuclear probability density in the donor-excited state, we use a horizontally polarized probe pulse of duration $\sigma_C = 0.25\,(2\pi/\omega)$ and center frequency $\Omega_C = \varepsilon_1 - 3\delta^2\,\omega$, which is resonant for $1 \to 0$ stimulated emission along the line $q_a = 2d$ and with $1' \to 0$ stimulated emission along the line $q_b = d$.[54] To preferentially detect the nuclear probability density in the acceptor-excited state, we use instead a vertically polarized probe, with duration $\sigma_C = 0.25\,(2\pi/\omega)$ and frequency $\Omega_C = \varepsilon_{1'} - 3\delta^2\,\omega$. This pulse is locally resonant with $1' \to 0$ stimulated emission near the line $q_b = 2d$.[55]

The pump-probe and pump-probe difference signals from an isotropic sample of the downhill dimer in the donor-detection case are plotted in Figure 9. The absence of even rough proportionality between the two signals attests to the effective influence of the control pulse on the time-course of short-time excitation transfer. A discernible

correspondence exists between the envelope of each of these signals and the corresponding donor-population trace in Fig. 8. But the hastening of excitation transfer during the first several vibrational periods by the action of the control pulse is not immediately obvious from the pump-probe difference signal.

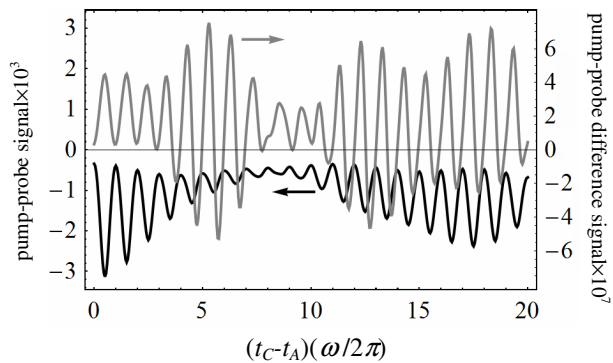

FIG. 9. Pump-probe signal (black curve) and pump-probe difference signal (gray curve) for isotropic sample of downhill EET dimer using VHH polarization scheme and probe-pulse center frequency which select primarily for nuclear probability density in the donor-excited state.

Interpretation of the pump-probe and pump-probe difference signals is complicated in the donor-detection case of Fig. 9 by the presence of significant excited-state absorption and some ground-state bleach, in addition to stimulated emission. These three contributions to both signals are shown separately in Figure 10. Stimulated emission (top panel) makes the largest contribution to both the pump-probe and pump-probe difference signals, and in both instances has a form similar to the overall signal. The excited-state absorption contribution to the pump-probe (pump-probe difference) signal has a maximum amplitude about 14% (49%) as large as stimulated emission, and is rather different in form. The ground-state bleach contribution to the pump-probe signal (bottom panel) is negligible, that to the pump-probe difference signal largely so. The relative sizes of these three contributions to the signals are consistent with the anticipated presence or absence of nuclear probability density in the vicinity of the line of local

22

resonance on the potential energy surface or surfaces from which each transition originates.[54]

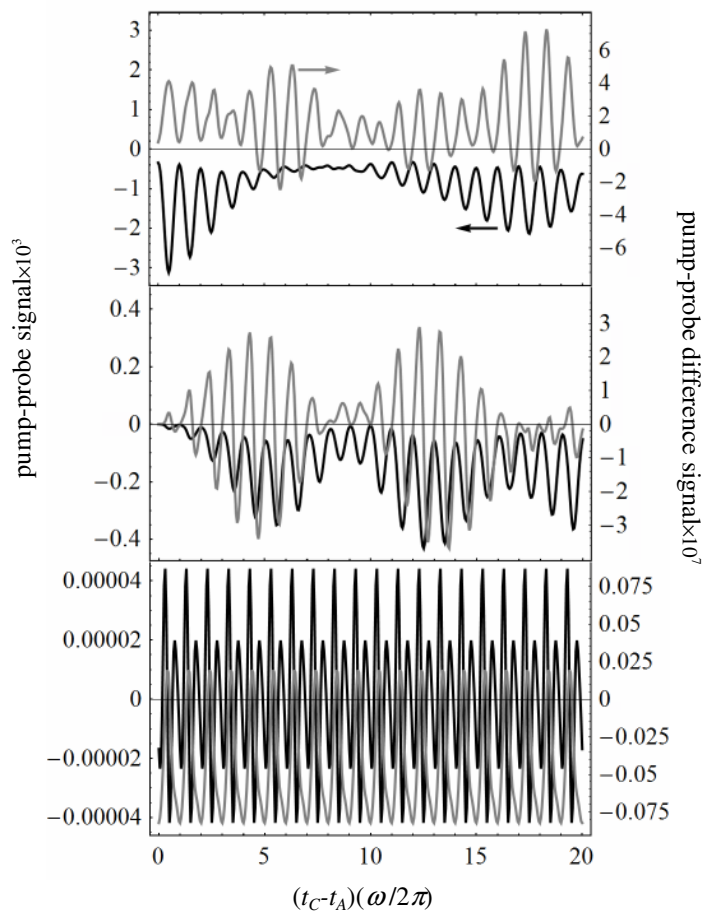

FIG. 10. Contributions from stimulated-emission (top panel), excited-state absorption (middle), and ground-state bleach (bottom) to pump-probe (black curves) and pump-probe difference (gray curves) to signals shown in Fig. 9.

The calculated pump-probe and pump-probe difference signals from an isotropic sample of the downhill dimer in the alternative, acceptor-detection case are shown in Figure 11. The strong influence of control-pulse-induced nuclear motion on the excitation-transfer process is again manifested by a lack of proportionality between the two signals. In this case, the envelopes of the two signals are very similar in form to the corresponding acceptor-state population (one minus the donor-state population plotted in

23

Fig. 8). Comparison of the stimulated-emission, excited-state absorption, and ground-state bleach contributions (not shown) shows that both the pump-probe and the pump-probe difference are entirely dominated by the first of these, as is consistent with arguments based on wave-packet access to the lines of local resonance.[55]

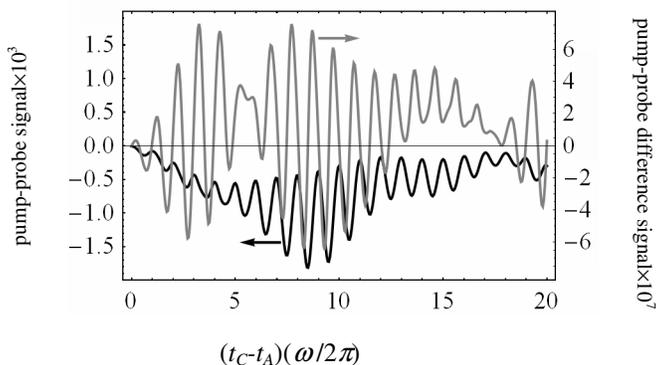

FIG. 11. Same as Fig. 9, but with VHV polarization and probe-pulse center frequency selecting for nuclear probability density in the acceptor-excited state.

## 3. DISCUSSION AND FUTURE PROSPECTS

The calculations of donor-state survival probability, shown in Section 2 for three different EET dimers, demonstrate that impulsive excitation of coherent vibrational motion in the acceptor chromophore can significantly alter the time-course of population transfer between donor- and acceptor-excited states. The pronounced effect of an externally driven acceptor vibration in impeding (or accelerating) short-time EET in the presence of moderately strong electronic-vibrational coupling in an equal-energy (or downhill) dimer is demonstrated in subsection 2A (2C). The diminished effect of a coherent vibration on electronic population transfer under weak electronic-vibrational coupling is illustrated for DTA-12 by the results of subsection 2B. These illustrative results provide the motivation for our more extensive investigation of the various corresponding spectroscopic pump-probe difference signals.

The at-best indirect correspondence between the nominally observable *total population* of the donor-excited state and the spectroscopic *signals* that would actually be measured in an ultrafast pump-probe or pump-probe difference experiment suggest that detailed system-specific calculations will generally be required to assist in the interpretation of future experiments along the lines proposed here and in Paper 1.

On the other hand, the calculations of subsection 2A demonstrate that the pump-probe difference signal (for appropriately chosen pulse center frequencies) can be dominated by its stimulated-emission component and that this signal component from an isotropic sample does not differ greatly from that for an oriented complex (which comprises just two wave-packet overlaps[45]). These findings suggest that similar simplifications should be possible in the calculation and analysis of the full nl-WPI difference signals for which the general expressions are given in the preceding paper. Domination of nl-WPI difference signals from EET systems by a small number of overlaps should greatly facilitate their interpretation in terms of the amplitude-level dynamics of surface-crossing wave packets.

The signal calculations presented here focus mainly on isolated model systems most closely representing energy-transfer dimers in the gas phase. Neglect of rotational motion is reasonable given the nanosecond-to-picosecond rotational periods estimated for molecules such as dithia-anthracenophane.[56, 57] But following the strategy of Ref. 10, the effects of medium-induced inhomogeneous broadening were mimicked in some cases by averaging calculations from EET systems having variable independent monomer site energies selected from a Gaussian distribution. Dynamical effects of embedding EET dimers in a liquid or crystalline environment could be simulated by combining Redfield[58,]

[59,60,61,62,63] or other relaxation theories with explicit dynamical treatment of strongly driven nuclear degrees of freedom, or perhaps by semiclassical wave-packet methods.[64] The full nl-WPI signals considered in Paper 1 depend on the preservation of electronic coherence over the relevant intrapulse-pair delays. But pump-probe difference measurements of impulsive vibrational control should be primarily sensitive to vibrational dephasing, and should therefore be applicable to condensed-phase systems on a timescale of tens of picoseconds.

The control strategy investigated here is motivated by the semiclassical notion that a Gaussian wave packet displaced from the $q_a = q_b$ line of intersection between donor- and acceptor-excited electronic states of equal site energy should be less susceptible to energy transfer than one sitting astride that line. As seen in subsections 2A, this strategy is effective under fairly strong electronic-vibrational coupling. On the other hand, subsection 2B shows that this simple displacement method is ineffective under weak electronic-vibrational coupling, where the spatial width of the wave packet is smaller than the excited-state shift in the equilibrium value of the vibrational coordinate, and the semiclassical picture of EET breaks down.

This failure under weak electronic-vibrational coupling does not, however, preclude the exertion of vibrational control over excitation transfer in such systems. With reference to Appendix A, we note, for instance, that the Franck-Condon overlap for DTA-12, $\left\langle (2_A)_{1'} \middle| (2_A)_1 \right\rangle = -0.03506$, happens to take a very small value. This small overlap should lead to slow excitation transfer from any state of the form $|1\rangle |\psi_1\rangle$ with

$$|\psi_1\rangle = U_S \otimes I_A \left| (0_S\, 2_A)_1 \right\rangle = U_S \otimes I_A \left\{ \tfrac{1}{2} \left| (2_a\, 0_b)_1 \right\rangle - \tfrac{1}{\sqrt{2}} \left| (1_a\, 1_b)_1 \right\rangle + \tfrac{1}{2} \left| (0_a\, 2_b)_1 \right\rangle \right\}.$$ [65] The donor-state survival probability following the preparation of such an initial state in DTA-12 is

26

plotted in Figure 12. For the example shown, we have imagined displacing the wave packet $|\psi_0\rangle = |(0_S\, 2_A)_0\rangle$ by $-d$ along $q_b$ prior to short-pulse absorption to the donor-excited state, so that $|\psi_1\rangle = e^{ip_b d}\,|(0_S 2_A)_0\rangle = e^{i\sqrt{2}P_S d}\,|(0_S\, 2_A)_1\rangle$. Although this illustration defers the question of how the initial state $|0\rangle|(0_S\, 2_A)_0\rangle$ could be prepared in practice, the greatly diminished initial decay of the donor-state population observed in Fig. 12 suggests that effective vibrational control over EET might well be possible even in the presence of weak electronic-vibrational coupling.

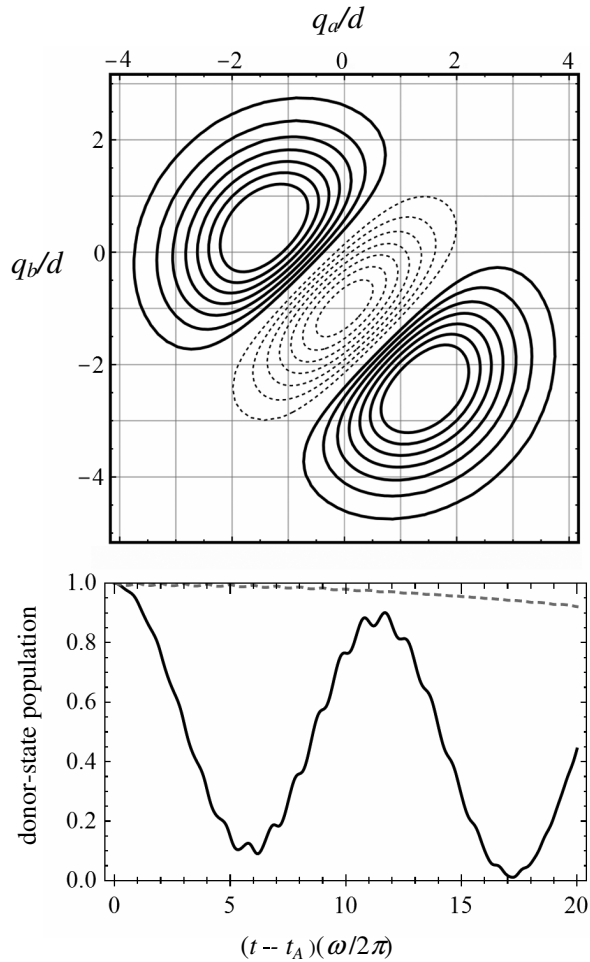

FIG. 12. Illustration of possible vibrational-control strategy for the DTA-12 model with its weak electronic-vibrational coupling. Upper panel shows spatially translated $\nu_A = 2$ vibrational wave function at instant of transfer to donor-excited state. Lower panel compares subsequent donor-state survival probability (gray

dashed line) to that following direct excitation from vibrational ground state (solid black line).

## ACKNOWLEDGEMENTS

## APPENDIX A: POPULATION TRANSFER FOR $J << \omega$

Oscillations at the vibrational frequency are generally less prominent in the survival probability traces of Figs. 1, 6, and 8 than they are in the corresponding pump-probe and pump-probe difference signals of Figs. 2, 5, and 9, respectively. While the latter signals are sensitive to the nuclear probability density in localized spatial windows, the former depend on the nuclear dynamics only through its effect on the detailed time-course of population transfer between the site-excited electronic states. When the EET coupling constant $J$ is small compared to the vibrational energy spacing, energy transfer occurs predominantly between vibronic states of nearly equal energy in the two site-state manifolds, if such energy matches exist. In the absence of interfering transition amplitudes from multiple vibronic levels of the donor, for instance, vibrational coherences do not impress themselves on the population evolution of a given vibronic level in the acceptor. The total acceptor population and the survival probability of the donor excitation are then largely devoid of vibrational-frequency oscillations.

Let us consider the electronic population dynamics in the one-exciton manifold of a system governed by

$$H_{one} = |1\rangle H_1 \langle 1| + |1'\rangle H_{1'} \langle 1'| + V \quad , \tag{17}$$

in which $H_1$ and $H_{1'}$ are given by Eqs. (7) and (8), respectively, with $\varepsilon_{1'} \leq \varepsilon_1$, and

$$V = J\left(\left|1'\right\rangle\left\langle1\right| + \left|1\right\rangle\left\langle1'\right|\right) \ . \tag{18}$$

The population evolution can be efficiently described by switching to symmetric and antisymmetric combinations of the donor and acceptor nuclear coordinates and momenta: $Q_S = (q_b + q_a)/\sqrt{2}$, $P_S = (p_b + p_a)/\sqrt{2}$, $Q_A = (q_b - q_a)/\sqrt{2}$, and $P_A = (p_b - p_a)/\sqrt{2}$ (see Figure 13). In terms of these operators,

$$H_1 = \frac{P_S^2 + P_A^2}{2m} + \frac{m\omega^2}{2}\left[\left(Q_S - \frac{d}{\sqrt{2}}\right)^2 + \left(Q_A + \frac{d}{\sqrt{2}}\right)^2\right] + \varepsilon_1 \tag{19}$$

and

$$H_{1'} = \frac{P_S^2 + P_A^2}{2m} + \frac{m\omega^2}{2}\left[\left(Q_S - \frac{d}{\sqrt{2}}\right)^2 + \left(Q_A - \frac{d}{\sqrt{2}}\right)^2\right] + \varepsilon_{1'} \ . \tag{20}$$

We denote by $\left|(v_S v_A)_1\right\rangle$ (or $\left|(v_S v_A)_{1'}\right\rangle$) an eigenket of $H_1$ (or $H_{1'}$) with energy $\omega(v_S + v_A + 1) + \varepsilon_1$ (or $\omega(v_S + v_A + 1) + \varepsilon_{1'}$). Note that the equilibrium values of $(Q_S, Q_A)$ are $(d/\sqrt{2}, -d/\sqrt{2})$ and $(d/\sqrt{2}, d/\sqrt{2})$ in state-1 and state-1′, respectively; only the equilibrium value of the antisymmetric coordinate differs in the two states.

We wish to determine the survival probability

$$P_1(t) = \left\langle\psi_1\left|\left\langle1\right|e^{iH_{one}t}\left|1\right\rangle\left\langle1\right|e^{-iH_{one}t}\left|1\right\rangle\right|\psi_1\right\rangle \tag{21}$$

for an initial state in the donor-excited manifold (with $\left\langle\psi_1\middle|\psi_1\right\rangle = 1$). This can be accomplished by solving the equation of motion

$$\frac{d}{dt}e^{i(H_{one}-V)t}e^{-iH_{one}t} = -i\tilde{V}(t)e^{i(H_{one}-V)t}e^{-iH_{one}t} \tag{22}$$

for the one-exciton time-evolution operator in the interaction picture, where

$$\tilde{V}(t) = e^{i(H_{one}-V)t}Ve^{-i(H_{one}-V)t} \ . \tag{23}$$

The interaction-picture "Hamiltonian" (23) has matrix elements

$$\left\langle (\nu_S' \nu_A')_{1'} \right| \left\langle 1' \left| \tilde{V}(t) \right| 1 \right\rangle \left| (\nu_S \nu_A)_1 \right\rangle = J e^{i(\varepsilon_{1'} - \varepsilon_1)t + i\omega(\nu_A' - \nu_A)t} \left\langle (\nu_A')_{1'} \left| (\nu_A)_1 \right\rangle \delta_{\nu_S', \nu_S} \right. \tag{24}$$

and

$$\left\langle (\nu_S \nu_A)_1 \right| \left\langle 1 \left| \tilde{V}(t) \right| 1' \right\rangle \left| (\nu_S' \nu_A')_{1'} \right\rangle = J e^{i(\varepsilon_1 - \varepsilon_{1'})t + i\omega(\nu_A - \nu_A')t} \left\langle (\nu_A)_1 \left| (\nu_A')_{1'} \right\rangle \delta_{\nu_S, \nu_S'} \right. \tag{25}$$

Elements (24) and (25) both oscillate at frequency $(\varepsilon_{1'} - \varepsilon_1) + \omega(\nu_A' - \nu_A)$; in the weak-coupling case $(J << \omega)$, they are therefore ineffectual unless $\varepsilon_{1'} + \nu_A' \omega \approx \varepsilon_1 + \nu_A \omega$.

In order to investigate the simplification that becomes possible when the site energies are equal or differ by multiple vibrational quanta, we assume $\varepsilon_1 - \varepsilon_{1'} = N\omega$. In this situation, it is reasonable to omit the time-dependent elements of $\tilde{V}$, and replace it by

$$\tilde{v} = J \sum_{\nu_S, \nu_A} \left| 1' \right\rangle \left| (\nu_S, \nu_A + N)_{1'} \right\rangle \left\langle (\nu_A + N)_{1'} \left| (\nu_A)_1 \right\rangle \left\langle (\nu_S, \nu_A)_1 \right| \left\langle 1 \right| + H.c. \tag{26} \right.$$

which is time-independent. Substitution in Eq. (22), integration, and return to the Schrödinger picture then yield

$$e^{-iH_{one}t} \cong e^{-i(H_{one} - V)t} e^{-i\tilde{v}t} . \tag{27}$$

Under this approximation, the survival probability (21) becomes

$$P_1(t) \cong \left\langle \psi_1 \right| \left\langle 1 \left| e^{i\tilde{v}t} \right| 1 \right\rangle e^{iH_1 t} e^{-iH_1 t} \left\langle 1 \left| e^{-i\tilde{v}t} \right| 1 \right\rangle \left| \psi_1 \right\rangle$$

$$= \left\langle \psi_1 \right| \left\langle 1 \left| \cos \tilde{v}t \right| 1 \right\rangle \left\langle 1 \left| \cos \tilde{v}t \right| 1 \right\rangle \left| \psi_1 \right\rangle$$

$$= \left\langle \psi_1 \right| \left\langle 1 \left| \cos^2 \tilde{v}t \right| 1 \right\rangle \left| \psi_1 \right\rangle$$

$$= \frac{1}{2} + \frac{1}{2} \sum_{\nu_S, \nu_A} \left| \left\langle (\nu_S \nu_A)_1 \left| \psi_1 \right\rangle \right|^2 \cos \left[ 2tJ \left\langle (\nu_A + N)_{1'} \left| (\nu_A)_1 \right\rangle \right] \right. \tag{28}$$

We want to evaluate Eq. (28) in situations where, as a result of impulsive control-pulse excitation, the initial wave packet takes the approximate form of the lowest state-0

vibrational eigenket displaced a distance $f$ along the acceptor mode (as illustrated in Fig. 13):

$$\left| \psi_1 \right\rangle = e^{-ip_b f} \left| (0_a 0_b)_0 \right\rangle$$

$$= e^{-iP_S f/\sqrt{2}} e^{-iP_A f/\sqrt{2}} \left| (0_S 0_A)_0 \right\rangle \ . \tag{29}$$

We also have

$$\left| (v_S v_A)_1 \right\rangle = e^{-ip_a d} \left| (v_S v_A)_0 \right\rangle$$

$$= e^{-iP_S d/\sqrt{2}} e^{iP_A d/\sqrt{2}} \left| (v_S v_A)_0 \right\rangle \ , \tag{30}$$

so the two-dimensional overlap in Eq. (28) becomes

$$\left\langle (v_S v_A)_1 \middle| \psi_1 \right\rangle = \frac{e^{-\frac{m\omega}{4}(f^2 + d^2)}}{\sqrt{v_S! v_A!}} \left[ \frac{\sqrt{m\omega}}{2}(f - d) \right]^{v_S} \left[ \frac{\sqrt{m\omega}}{2}(f + d) \right]^{v_A} \ . \tag{31}$$

The survival probability is thereby obtained in the form

$$P_1(t) = \frac{1}{2} + \frac{1}{2} e^{-\frac{m\omega}{4}(f+d)^2} \sum_{v_A=0}^{\infty} \frac{1}{v_A!} \left[ \frac{m\omega}{4}(f+d)^2 \right]^{v_A} \cos\left[ 2tJ \left\langle (v_A + N)_{1'} \middle| (v_A)_1 \right\rangle \right] \ . \tag{32}$$

The requisite one-dimensional Franck-Condon overlaps are straightforwardly evaluated using the relation

$$\left\langle (v_A + N)_{1'} \middle| (v_A)_1 \right\rangle = (-\sqrt{2}\,\delta)^N e^{-\delta^2} \sum_{k=0}^{v_A} (-2\delta^2)^k \frac{\sqrt{(v_A + N)! v_A!}}{k!(k+N)!(v_A - k)!} \ . \tag{33}$$

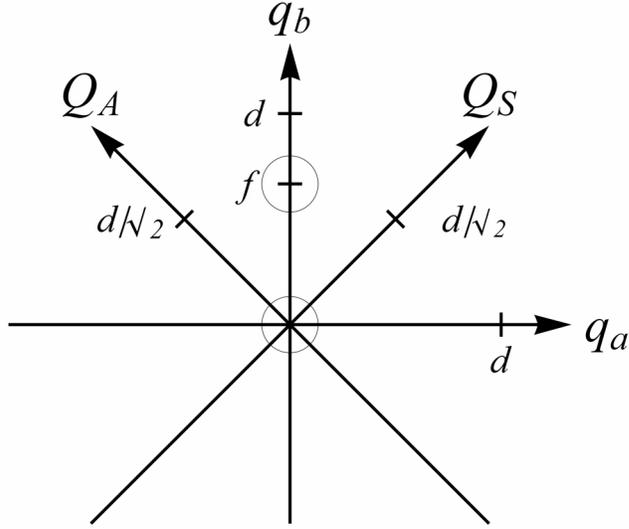

FIG. 13. Symmetric and antisymmetric vibrational coordinates plotted on donor and acceptor coordinate axes. The lowest-energy vibrational state in the donor-excited [acceptor-excited] electronic state is centered at $(q_a, q_b) = (d, 0)$ [ $(q_a, q_b) = (0, d)$ ]. Also shown are the location (at the origin) of the lowest vibrational state in the electronic ground state and the location of a ground-state wave packet displaced by $f$ along the $q_b$-axis.

In the case of "ideal" displacement $f = -d$, the approximate survival probability (32) reduces to

$$P_1(t) = \frac{1}{2} + \frac{1}{2} \cos\left[ 2tJ \left\langle (N_A)_{1'} \big| (0_A)_1 \right\rangle \right] ,$$ (34)

and population transfer proceeds as if the state-1 wave packet were the lowest vibrational level of the donor-excited electronic state. This behavior is to be compared with the survival probability following direct Franck-Condon excitation $(f = 0)$,

$$P_1(t) = \frac{1}{2} + \frac{1}{2} e^{-\delta^2/2} \sum_{\nu_A=0}^{\infty} \frac{(\delta^2/2)^{\nu_A}}{\nu_A!} \cos\left[ 2tJ \left\langle (\nu_A + N)_{1'} \big| (\nu_A)_1 \right\rangle \right] ,$$ (35)

in which non-zero populations in higher donor-excited levels also contribute. The relative speed of short-time EET evidently hinges on the size of the overlaps (33) for contributing donor-excited vibronic states.

The weak electronic-vibrational coupling limit ($\delta^2 \ll 1$) is exemplified for the case of equal site energies by the DTA-12 calculations of subsection 2B. In this system, the vibrationally excited antisymmetric-mode overlaps are all smaller in size than $\langle (0_A)_{1'} | (0_A)_1 \rangle$, so the short-time donor-population falls off *more rapidly* following acceptor-mode displacement (Eq. (32) with $f = -0.54d$) than it does under Franck-Condon excitation (Eq. (35)). These predictions of the weak EET-coupling approximation are consistent with the rigorously calculated population dynamics shown in Fig. 6. In addition, they differ only slightly from the survival probability calculated by means of Eq. (28) using the $|\psi_1\rangle$ actually generated by the sequential action of the control and pump pulses, rather than the approximate form (29) (results not shown).

The model system studied in subsection 2A features equal site energies, moderate electronic-vibrational coupling ($\delta^2 = 2.5$), and moderate energy-transfer coupling ($J = 0.2\omega$). For this system under "ideal" impulsive displacement of the acceptor mode ($f = -d$), Eq. (34) with $N_A = 0$ predicts

$$P_1(t) = \frac{1}{2} + \frac{1}{2}\cos\left(\frac{2Jt}{e^{5/2}}\right) . \tag{36}$$

According to Eq. (35), vertical donor excitation without prior vibrational displacement leads to

$$P_1(t) \cong \frac{1}{2} + \frac{1}{2e^{5/4}}\left[\cos\left(\frac{2Jt}{e^{5/2}}\right) + \frac{5}{4}\cos\left(\frac{8Jt}{e^{5/2}}\right) + \frac{25}{32}\cos\left(\frac{7Jt}{e^{5/2}}\right)\right.$$

33

$$+\frac{125}{384}\cos\left(\frac{16Jt}{3e^{5/2}}\right)+\frac{625}{6144}\cos\left(\frac{31Jt}{12e^{5/2}}\right)+\cdots\right] . \tag{37}$$

Here $\nu_A = 1-4$ all have larger overlaps than $\nu_A = 0$, so the donor-state population falls off more rapidly after direct Franck-Condon excitation than it does following prior acceptor-mode displacement. For the displacement, $f = -0.338d$, achieved by the impulsive Raman process described in subsection 2A, Eq. (32) predicts

$$P_1(t) \cong \frac{1}{2}+\frac{1}{2e^{0.5478}}\left[\cos\left(\frac{2Jt}{e^{5/2}}\right)+0.5478\cos\left(\frac{8Jt}{e^{5/2}}\right)+0.1500\cos\left(\frac{7Jt}{e^{5/2}}\right)\right.$$

$$\left.+0.0274\cos\left(\frac{16Jt}{3e^{5/2}}\right)+0.0038\cos\left(\frac{31Jt}{12e^{5/2}}\right)+\cdots\right] . \tag{38}$$

This function is plotted along with the rigorously calculated population in Fig. 1; a similar form that agrees slightly better with the rigorous result (not shown) is obtained from Eq. (28) evaluated with the numerically calculated components $\langle(\nu_S\nu_A)_1|\psi_1\rangle$ generated by the actual impulsive Raman process and subsequent pump-pulse excitation.

In the intermediate electronic-vibrational coupling case $\delta = 1$, the Franck-Condon overlaps for $N = 0$ given by Eq. (33) depend only weakly (in magnitude) on the antisymmetric-mode quantum number, taking values $1/e$, $-1/e$, $-1/e$, $-1/3e$, and $1/3e$ for $\nu_A = 0$ through 4, respectively. These overlaps give rise to survival probabilities

$$P_1(t) = \frac{1}{2}+\frac{1}{2}\cos\left(\frac{2Jt}{e}\right) \tag{39}$$

for $f = -d$ and

$$P_1(t) \cong \frac{1}{2}+\frac{13}{16e^{1/2}}\cos\left(\frac{2Jt}{e}\right)+\frac{3}{256e^{1/2}}\cos\left(\frac{2Jt}{3e}\right)$$

$$=\frac{1}{2}+0.493\cos\left(\frac{2Jt}{e}\right)+0.007\cos\left(\frac{2Jt}{3e}\right) \tag{40}$$

34

for $f = 0$, so the acceptor-mode displacement has only a small effect in this case. This pattern suggests that simple displacement of a moderately Franck-Condon active acceptor vibration, such as mode-6 of anthracene (with $\delta_6 \cong 1.05$   $\omega_6 \cong 1400 \, \mathrm{cm}^{-1}$)[66] in DTA, would be largely ineffective as a means of controlling electronic excitation transfer. We have verified this prediction with numerical calculations (not shown) of the donor-state survival probability and pump-probe signal for a DTA-6 model analogous to the DTA-12 results of subsection 2B.

In the downhill model considered in subsection 2C, the relevant overlaps are $\left\langle (v_A + 5)_{1'} \middle| (v_A)_1 \right\rangle =$ -0.4189, -0.1710, 0.1371, 0.2706, and 0.2037 for $v_A = 0 - 4$, respectively. The corresponding population-weights to be used in Eq. (32) with $f = 0$ and $f = -0.338d$ are the same as those appearing in Eqs. (37) and (38), respectively. The resulting, approximate predictions for $P_1(t)$ are plotted in Fig. 8 along with the rigorously calculated traces.

allowing experimental discrimination between emission signals from the 1- and 2-exciton manifolds.

[41] Y. J. Yan and S. Mukamel, "Femtosecond pump-probe spectroscopy of polyatomic molecules in condensed phases," Phys. Rev. A **41**, 6485 (1990).

[42] D. Egorova, M. F. Gelin, and W. Domcke, "Time- and frequency-resolved fluorescence spectra of nonadiabatic dissipative systems: What photons can tell us," J. Chem. Phys. **122**, 134504 (2005).

[43] Qualitative agreement between the envelope of the predicted signal and site population dynamics—to be regarded as an abstract "observable," rather than a quantity readily available for spectroscopic determination on the femtosecond timescale—may not obtain in general. Calculations of populations and signals from similar model systems with stronger electronic-vibrational coupling (larger $\delta$, results not shown) exhibited less correspondence between the pump-probe envelope and the donor-excitation survival probability. In the case of stronger electronic-vibrational coupling, the various electronic difference potentials are steeper, and the nuclear probability density in the correspondingly narrowed probe-window regions may become less reflective of the site-state population as a whole.

[44] $q_a = 2d$ and $q_b = 2d$ are the lines of $1 \rightarrow 0$ and $1' \rightarrow 0$ stimulated-emission resonance, respectively; both lines are visited by significant nuclear probability density in the appropriate singly-excited state. $q_a = 2d$ is also the resonance line for $1 \leftarrow 0$ ground-state bleach and $2 \leftarrow 1'$ excited-state absorption, while $q_b = 2d$ is the line for $1' \leftarrow 0$ ground-state bleach and $2 \leftarrow 1$ excited-state absorption; but these regions are

never populated by significant nuclear probability density in the electronic state from which the transition would take place.

[45] The stimulated-emission pump-probe difference signal plotted in Fig. 3 is given by $\Theta_1^{SE} + \Theta_{1'}^{SE}$, where

$$
\begin{aligned}
\Theta_\varepsilon^{SE} = 8\,\text{Re}\,\frac{3}{35}\big\{ & \big\langle \{a(1'0)\}_\varepsilon \,\big|\, \{c(1'0)c(01)a(1'0)p(01)p(10)\}_\varepsilon \big\rangle \\
+ & \big\langle \{a(10)\}_\varepsilon \,\big|\, \{c(10)c(01)a(10)p(01')p(1'0)\}_\varepsilon \big\rangle \\
+ & \big\langle \{a(1'0)p(01)p(10)\}_\varepsilon \,\big|\, \{c(1'0)c(01)a(1'0)\}_\varepsilon \big\rangle \\
+ & \big\langle \{a(10)p(01')p(1'0)\}_\varepsilon \,\big|\, \{c(10)c(01)a(10)\}_\varepsilon \big\rangle \big\}
\end{aligned}
$$

(see Appendix C (online) and Table 1 in Appendix A of Paper 1). Because no further propagation need be carried out after the reduced $C$-pulse propagator acts, only two of the four terms in this equation can be nonzero for either choice of $\varepsilon$ (i.e. the second and fourth (first and third) terms for $\varepsilon = 1\,(1')$ ).

[46] The vertical scales are the same in Fig. 2 and the upper panel of Fig. 4. The signals are larger in the latter case because of the longer pump and probe pulses.

[47] See EPAPS supplementary material at [URL will be inserted by AIP] for Appendix C of this document.

[48] I. Yamazaki, S. Akimoto, T. Yamazaki, S. Sato, and Y. Sakata, "Oscillatory excitation transfer in dithiaanthracenophane: Quantum beat in a coherent photochemical process in solution," J. Phys. Chem. A **106,** 2122 (2002).

[49] I. Yamazaki, S. Akimoto, N. Aratani, and A. Osuka, "Observation of coherent recurrence motion of excitons in anthracene dimers," Bull. Chem. Soc. Jpn. **77,** 1959 (2004).

[54] When $\Omega_C = \varepsilon_1 - 3\delta^2\,\omega$, the spatial band $q_a = 2d$ is also the locus of resonant $1 \leftarrow 0$ ground-state bleach and $2 \leftarrow 1'$ excited-state absorption. This region is largely inaccessible in the electronic ground state, so ground-state bleaching should be very slight. It may be accessible to some degree following EET to the acceptor-excited state, especially from a wave packet displaced along the acceptor mode by control-induced ISRS before excitation to the donor-state. The band along $q_b = d$ is also the locus of $1' \leftarrow 0$ ground-state bleach and $2 \leftarrow 1$ excited-state absorption with this same probe frequency. The former process should again be weak, but the latter may occur to some degree in the presence of control-induced acceptor-mode displacement, despite being discriminated against by the horizontally polarized probe in the VHH scheme. These

qualitative predictions are consistent with the significant signal contributions from excited-state absorption shown in the middle panel of Fig. 10.

[55] With $\Omega_C = \varepsilon_{1'} - 3\delta^2 \omega$, $q_b = 2d$ is also the line of resonance for $1' \leftarrow 0$ ground-state bleach and $2 \leftarrow 1$ excited-state absorption, both of which should be negligible. $q_a = 3d$ is the line of resonance for $1 \rightarrow 0$ stimulated emission, $1 \leftarrow 0$ ground-state bleach, and $2 \leftarrow 1'$ excited-state absorption with this probe frequency, all which are expected to be essentially undetectable.

[56] J. D. Biggs, unpublished.

[57] For a recent demonstration of coherent control and state reconstruction of rotational wave packets, see H. Hasegawa and Y. Ohshima, "Quantum state reconstruction of a rotational wave packet created by a nonresonant intense femtosecond laser field," Phys. Rev. Lett. **101**, 053002 (2008).

[58] W. T. Pollard, A. K. Felts, and R. A. Friesner, "The Redfield equations in condensed-phase quantum dynamics," Adv. Chem. Phys. **93**, 77 (1996).

[59] A. Matro and J. A. Cina, "Theoretical study of time-resolved fluorescence anisotropy from coupled chromophore pairs," J. Phys. Chem. **99**, 2568 (1995).

[60] L. W. Ungar and J. A. Cina, "The relaxation dynamics and short-time optical response of a multimode open system," J. Phys. Chem. A **102**, 7382 (1998).

[61] D. Egorova and W. Domcke, "Coherent vibrational dynamics during ultrafast photoinduced electron-transfer reactions: quantum dynamical simulations within multilevel Redfield theory," Chem. Phys. Lett. **384**, 157 (2004).

[62] M. Yang and G. R. Fleming, "Influence of phonons on exciton transfer dynamics: comparison of the Redfield, Förster, and modified Redfield equations," Chem. Phys.



\noindent{\emph{Online supplement to:``Calculations of Nonlinear Wave-Packet Interferometry Signals in the Pump-Probe Limit as Tests for Vibrational Control over Electronic Excitation Transfer,'' J.D. Biggs and J.A. Cina. }}

\

\appendix
\setcounter{section}{2}

\noindent{\large\bfseries \fontfamily{phv}\selectfont
\MakeUppercase{Appendix B: VIBRONIC MATRIX ELEMENTS FOR SECOND-ORDER PULSE PROPAGATORS}}

\

In going to the pump-probe limit of a nl-WPI experiment, the interpulse delays for pulses within a phase-related pair are set to zero.  The treatment described in Paper 1 (see Eq. (13)) properly accounts for the non-negligible effects of pulse overlap that enter in this limit.  We construct second-order pulse propagators for ground-state bleach (GSB), stimulated emission (SE), or excited-state absorption (ESA) depending on the particular action of the pulse ($0 \to \varepsilon  \to 0$
, $\varepsilon  \to 0 \to \varepsilon $
,  or $\varepsilon  \to 2 \to \varepsilon $
 respectively).
	
Propagators for GSB take the form
\begin{equation}\begin{gathered}
  p_I^{(01)} (\infty ;\,t_2 )p_I^{(10)} (t_2 ;\,t_1 ) = \left( {\frac{{iE_I m}}
{2}} \right)^2 \int\limits_{ - {\kern 1pt} \infty }^\infty  {dt_2 } \int\limits_{ - {\kern 1pt} \infty }^{t_2 } {dt_1 \,} f_I (t_2 )f_I (t_1 )e^{i\Phi _I (t_2 ) - i\Phi _I (t_1 )}  \hfill \\
  \quad \quad \quad \quad \quad \quad \quad \quad \quad \quad \;\;\,\,\,\,\,\,\,\,\,\,\,\,\,\,\,\,\,\,\,\,\,\,\,\,\,\,\,\,\,\,\,\, \times e^{ - iH_0 (t_I  - t_2 )} e^{ - iH_1 (t_2  - t_I )} e^{ - iH_1 (t_I  - t_1 )} e^{ - iH_0 (t_1  - t_I )} \;\,. \hfill \\
\end{gathered}\end{equation}
To calculate the matrix elements for this operator in the eigenbasis of $H_0 $, we introduce completeness relations in the eigenbasis of $H_1 $,
\begin{equation}
\begin{gathered}
  \left\langle \left( {\nu _a ,\nu _b } \right)_0 \left| {p_I^{(01)} (\infty ;\,t_2 )p_I^{(10)} (t_2 ;\,t_1 )} \right| {\left( {\bar \nu _a ,\bar \nu _b } \right)_0 } \right\rangle  \hfill \\
  \quad \quad \quad \quad \quad \quad \quad  = \left( {\frac{{iE_I m}}
{2}} \right)^2 \int\limits_{ - {\kern 1pt} \infty }^\infty  {dt_2 } \int\limits_{ - {\kern 1pt} \infty }^{t_2 } {dt_1 \,} e^{ - {{(t_1  - t_I )^2 } \mathord{\left/
 {\vphantom {{(t_1  - t_I )^2 } {2\sigma _I^2 }}} \right.
 \kern-\nulldelimiterspace} {2\sigma _I^2 }}} e^{ - {{(t_2  - t_I )^2 } \mathord{\left/
 {\vphantom {{(t_2  - t_I )^2 } {2\sigma _I^2 }}} \right.
 \kern-\nulldelimiterspace} {2\sigma _I^2 }}} e^{i\Omega _I (t_2  - t_I ) - i\Omega _I (t_1  - t_I )}  \hfill \\
  \,\,\,\,\,\,\,\,\,\,\,\,\,\,\,\,\,\,\,\,\,\,\,\,\,\,\,\,\,\,\,\,\,\,\,\,\,\,\,\,\,\,\,\,\,\,\,\, \times \left\langle \left( {\nu _a ,\nu _b } \right)_0 \right| {e^{ - iH_0 (t_I  - t_2 )} e^{ - iH_1 (t_2  - t_I )} e^{ - iH_1 (t_I  - t_1 )} e^{ - iH_0 (t_1  - t_I )} } \left| {\left( {\bar \nu _a ,\bar \nu _b } \right)_0 } \right\rangle  \hfill \\
  \,\,\,\,\,\,\,\,\,\,\,\,\,\,\,\,\,\,\,\,\,\,\,\,\,\,\,\,\,\,\,\,\,\,\,\,\,\,\,\,\,\, = \delta _{\nu _b , \bar \nu _b } \left( {\frac{{iE_I m}}
{2}} \right)^2 \sum\limits_{\bar{\bar \nu} _a } {\left\langle {(\nu _a )_g } \right|} \left. {(\,\bar{\bar \nu} _a )_e } \right\rangle \left\langle {(\,\bar{\bar \nu} _a )_e } \right|\left. {(\,\bar \nu _a )_g } \right\rangle  \hfill \\
  \quad \quad \quad \quad \quad \quad {\kern 1pt} \,\,\,\,\,\,\,\,\,\,\,\,\, \times \int\limits_{ - {\kern 1pt} \infty }^\infty  {d\tau _2 } \int\limits_{ - {\kern 1pt} \infty }^{\tau _2 } {d\tau _1 \,} \exp \left\{ { - \frac{{\tau _1^2 }}
{{2\sigma _I^2 }} + i\,(\omega (\bar{\bar \nu} _a  - \bar \nu _a ) + \varepsilon _1  - \Omega _I )\,\tau _1 } \right\} \hfill \\
  \quad \quad \quad \quad \quad \quad \;\,\,\,\,{\kern 1pt} \,\,\,\,\,\,\,\,\,\,\,\,\,\,\,\,\,\,\,\,\,\,\,\,\,\,\,\,\,\,\,\,\,\,\,\,\,\,\,\, \times \exp \left\{ { - \frac{{\tau _2^2 }}
{{2\sigma _I^2 }} - i\,(\omega (\bar{\bar \nu} _a  - \nu _a ) + \varepsilon _1  - \Omega _I )\,\tau _2 } \right\}\;\,. \hfill \\
\end{gathered}
\end{equation}
A closed-form expression can be derived for nested time integrals of the form appearing in Eq. (B.2), due to the Gaussian nature of the integrand.  The first step is to change the order of integration,
	\begin{equation}
\begin{aligned}
  I(\alpha ,\beta ) &\equiv \int\limits_{ - {\kern 1pt} \infty }^\infty  {d\tau _2 } \int\limits_{ - {\kern 1pt} \infty }^{\tau _2 } {d\tau _1 \,} \exp \left\{ { - \frac{{\tau _1^2 }}
{{2\sigma _I^2 }} + i\,\alpha \,\tau _1  - \frac{{\tau _2^2 }}
{{2\sigma _I^2 }} - i\,\beta \,\tau _2 } \right\} \\
&= \int\limits_{ - {\kern 1pt} \infty }^\infty  {d\tau _1 } \int\limits_{\tau _1 }^\infty  {d\tau _2 \,} \exp \left\{ { - \frac{{\tau _1^2 }}
{{2\sigma _I^2 }} + i\,\alpha \,\tau _1  - \frac{{\tau _2^2 }}
{{2\sigma _I^2 }} - i\,\beta \,\tau _2 } \right\}\,\,. \\
\end{aligned}
\end{equation}
We now make a change of variables, letting $\tau  = \tau _2  - \tau _1 $
\begin{equation}
I(\alpha ,\beta ) = \int\limits_{ - {\kern 1pt} \infty }^\infty  {d\tau _1 } \int\limits_0^\infty  {d\tau \,} \,\exp \left\{ { - \frac{{\tau _1^2 }}
{{2\sigma _I^2 }} + i\,\alpha \,\tau _1  - \frac{{\tau _1^2  + \tau ^2  + 2\,\tau \,\tau _1 }}
{{2\sigma _I^2 }} - i\,\beta \,(\tau  + \tau _1 )} \right\}\,,
\end{equation}
and then change integration order again to obtain
	\begin{equation}
I(\alpha ,\beta ) = \int\limits_0^\infty  {d\tau \exp } \left\{ { - \frac{{\tau ^2 }}
{{2\sigma _I^2 }} - i\,\beta \,\tau } \right\}\int\limits_{ - {\kern 1pt} \infty }^\infty  {d\tau _1 } \,\exp \left\{ { - \frac{{\tau _2^2 }}
{{\sigma _I^2 }} + i\,(\alpha  - \beta )\tau _1  - \frac{{\tau \,\tau _1 }}
{{\sigma _I^2 }}} \right\}\,.
\end{equation}
The inner integral is solved in the usual way, giving
	\begin{equation}
\begin{aligned}
  I(\alpha ,\beta ) &= \sqrt \pi  \sigma \exp \left\{ { - \frac{{(\alpha  - \beta )^2 }}
{{4\sigma _I^2 }}} \right\}\int\limits_0^\infty  {d\tau \exp } \left\{ { - \frac{{\tau ^2 }}
{{2\sigma _I^2 }} - i\,\beta \,\tau  + \frac{{\sigma _I^2 }}
{4}\left( {i\,(\alpha  - \beta ) - \frac{{\tau \,}}
{{\sigma _I^2 }}} \right)^2 } \right\}\, \hfill \\
  &= \sqrt \pi  \sigma \exp \left\{ { - \frac{{(\alpha  - \beta )^2 }}
{{4\sigma _I^2 }}} \right\}\int\limits_0^\infty  {d\tau \exp } \left\{ { - \frac{{\tau ^2 }}
{{4\sigma _I^2 }} - i\,\tau \left( {\frac{{\alpha  + \beta }}
{2}} \right)} \right\}\,\,\,, \hfill \\
\end{aligned}
\end{equation}
which can be written in terms of the error function,
	\begin{equation}
I(\alpha ,\beta ) = \pi \,\sigma _I^2 \exp \left\{ { - \frac{{\sigma _I^2 }}
{2}(\alpha ^2  + \beta ^2 )} \right\}\left( {1 - {\text{erf}}\left\{ {\frac{{i\sigma _I }}
{2}(\alpha  + \beta )} \right\}} \right).
\end{equation}
We can now return to the vibronic matrix element for the GSB propagator, which by Eq. (B.7) can be written as
	\begin{equation}
\begin{gathered}
  \left\langle \left( {\nu _a ,\nu _b } \right)_0 \right| {p_I^{(01)} (\infty ;\,t_2 )p_I^{(10)} (t_2 ;\,t_1 )} \left| {\left( {\bar \nu _a ,\bar \nu _b } \right)_0 } \right\rangle  \hfill \\
  \quad \quad \quad \quad \quad \quad \quad  = \delta _{\nu _b , \bar \nu _b } \left( {\frac{{iE_I m}}
{2}} \right)^2 \sum\limits_{\bar{\bar \nu} _a } {\left\langle {(\nu _a )_g } \right|} \left. {(\,\bar{\bar \nu} _a )_e } \right\rangle \left\langle {(\,\bar{\bar \nu} _a )_e } \right|\left. {(\,\bar \nu _a )_g } \right\rangle  \hfill \\
  \quad \quad \quad \quad \quad \quad \;\,\,\,\,{\kern 1pt} \,\,\,\,\,\,\,\,\,\,\,\,\, \times \,I\left( {\omega (\bar{\bar \nu} _a  - \bar \nu _a ) + \varepsilon _1  - \Omega _I ,\omega (\bar{\bar \nu} _a  - \nu _a ) + \varepsilon _1  - \Omega _I } \right)\,\,\,. \hfill \\
\end{gathered}
\end{equation}
Matrix elements of the operator $
p_I^{(01')} (\infty ;\,t_2 )p_I^{(1'0)} (t_2 ;\,t_1 )
$
 follow analogously.  A GSB propagator also describes the action of the pre-resonant control pulse, whose effect on the system is primarily the creation of a ground-state wave packet, and Eq. (15) of the main text is equivalent to Eq. (B.8).

The ESA propagators are obtained in a similar fashion,
	\begin{equation}
\begin{gathered}
  \left\langle \left( {\nu _a ,\nu _b } \right)_1 \right| {p_I^{(12)} (\infty ;\,t_2 )p_I^{(21)} (t_2 ;\,t_1 )} \left| {\left( {\bar \nu _a ,\bar \nu _b } \right)_1 } \right\rangle  \hfill \\
  \quad \quad \quad \quad \quad \quad \quad  = \left( {\frac{{iE_I m}}
{2}} \right)^2 \int\limits_{ - {\kern 1pt} \infty }^\infty  {dt_2 } \int\limits_{ - {\kern 1pt} \infty }^{t_2 } {dt_1 \,} e^{ - {{(t_1  - t_I )^2 } \mathord{\left/
 {\vphantom {{(t_1  - t_I )^2 } {2\sigma _1^2 }}} \right.
 \kern-\nulldelimiterspace} {2\sigma _1^2 }}} e^{ - {{(t_2  - t_I )^2 } \mathord{\left/
 {\vphantom {{(t_2  - t_I )^2 } {2\sigma _1^2 }}} \right.
 \kern-\nulldelimiterspace} {2\sigma _1^2 }}} e^{i\Omega _I (t_2  - t_I ) - i\Omega _I (t_1  - t_I )}  \hfill \\
  \,\,\,\,\,\,\,\,\,\,\,\,\,\,\,\,\,\,\,\,\,\,\,\,\,\,\,\,\,\,\,\,\,\,\,\,\,\,\,\,\,\,\,\,\,\,\,\, \times \left\langle \left( {\nu _a ,\nu _b } \right)_1 \right| {e^{ - iH_1 (t_I  - t_2 )} e^{ - iH_2 (t_2  - t_I )} e^{ - iH_2 (t_I  - t_1 )} e^{ - iH_1 (t_1  - t_I )} } \left| {\left( {\bar \nu _a ,\bar \nu _b } \right)_1 } \right\rangle \, \hfill \\
  \,\,\,\,\,\,\,\,\,\,\,\,\,\,\,\,\,\,\,\,\,\,\,\,\,\,\,\,\,\,\,\,\,\,\,\,\,\,\,\,\,\, = \delta _{\nu _a , \bar \nu _a } \left( {\frac{{iE_I m}}
{2}} \right)^2 \sum\limits_{\bar{\bar \nu} _b } {\left\langle {(\nu _b )_g } \right|} \left. {(\,\bar{\bar \nu} _b )_e } \right\rangle \left\langle {(\,\bar{\bar \nu} _b )_e } \right|\left. {(\,\bar \nu _b )_g } \right\rangle  \hfill \\
  \quad \quad \quad \quad \quad \quad \;\,\,\,\,{\kern 1pt} \,\,\,\,\,\,\,\,\,\,\,\,\, \times I\left( {\omega (\bar{\bar \nu} _b  - \bar \nu _b ) + (\varepsilon _2  - \varepsilon _1 ) - \Omega _I ,\omega (\bar{\bar \nu} _b  - \nu _b ) + (\varepsilon _2  - \varepsilon _1 ) - \Omega _I } \right)\;, \hfill \\
\end{gathered}
\end{equation}
and the corresponding elements of $
p_I^{(1'2)} (\infty ;\,t_2 )p_I^{(21')} (t_2 ;\,t_1 )
$ follow by direct analogy. Unlike in the case of GSB, we now have the possibility that the final and initial electronic states will be different.  We then need matrix elements of the form
	\begin{equation}
\begin{gathered}
  \left\langle \left( {\nu _a ,\nu _b } \right)_{1'} \right| {p_I^{(1'2)} (\infty ;\,t_2 )p_I^{(21)} (t_2 ;\,t_1 )} \left| {\left( {\bar \nu _a ,\bar \nu _b } \right)_1 } \right\rangle  \hfill \\
  \quad \quad \quad \quad \quad \quad \quad  = \left( {\frac{{iE_I m}}
{2}} \right)^2 \int\limits_{ - {\kern 1pt} \infty }^\infty  {dt_2 } \int\limits_{ - {\kern 1pt} \infty }^{t_2 } {dt_1 \,} e^{ - {{(t_1  - t_I )^2 } \mathord{\left/
 {\vphantom {{(t_1  - t_I )^2 } {2\sigma _1^2 }}} \right.
 \kern-\nulldelimiterspace} {2\sigma _1^2 }}} e^{ - {{(t_2  - t_I )^2 } \mathord{\left/
 {\vphantom {{(t_2  - t_I )^2 } {2\sigma _1^2 }}} \right.
 \kern-\nulldelimiterspace} {2\sigma _1^2 }}} e^{i\Omega _I (t_2  - t_I ) - i\Omega _I (t_1  - t_I )}  \hfill \\
  \,\,\,\,\,\,\,\,\,\,\,\,\,\,\,\,\,\,\,\,\,\,\,\,\,\,\,\,\,\,\,\,\,\,\,\,\,\,\,\,\,\,\,\,\,\,\,\, \times \left\langle \left( {\nu _a ,\nu _b } \right)_1 \right| {e^{ - iH_{1'} (t_I  - t_2 )} e^{ - iH_2 (t_2  - t_I )} e^{ - iH_2 (t_I  - t_1 )} e^{ - iH_1 (t_1  - t_I )} } \left| {\left( {\bar \nu _a ,\bar \nu _b } \right)_1 } \right\rangle \, \hfill \\
  \,\,\,\,\,\,\,\,\,\,\,\,\,\,\,\,\,\,\,\,\,\,\,\,\,\,\,\,\,\,\,\,\,\,\,\,\,\,\,\,\,\, = \left( {\frac{{iE_I m}}
{2}} \right)^2 \left\langle {{(\nu _a )_g }}
 \mathrel{\left | {\vphantom {{(\nu _a )_g } {(\,\bar \nu _a )_e }}}
 \right. \kern-\nulldelimiterspace}
 {{(\,\bar \nu _a )_e }} \right\rangle \left\langle {(\,\nu _b )_e } \right|\left. {(\,\bar \nu _b )_g } \right\rangle  \hfill \\
  \quad \quad \quad \quad \quad \quad \;\,\,\,\,{\kern 1pt} \,\,\,\,\,\,\,\,\,\,\,\,\, \times I\left( {\omega (\nu _b  - \bar \nu _b ) + (\varepsilon _2  - \varepsilon _1 ) - \Omega _I ,\omega (\bar \nu _a  - \nu _a ) + (\varepsilon _2  - \varepsilon _{1'} ) - \Omega _I } \right)\;. \hfill \\
\end{gathered}
\end{equation}
Finally, the SE propagators are written
	\begin{equation}
\begin{gathered}
  \left\langle \left( {\nu _a ,\nu _b } \right)_1 \right| {p_I^{(10)} (\infty ;\,t_2 )p_I^{(01)} (t_2 ;\,t_1 )} \left| {\left( {\bar \nu _a ,\bar \nu _b } \right)_1 } \right\rangle  \hfill \\
  \quad \quad \quad \quad \quad \quad \quad  = \left( {\frac{{iE_I m}}
{2}} \right)^2 \int\limits_{ - {\kern 1pt} \infty }^\infty  {dt_2 } \int\limits_{ - {\kern 1pt} \infty }^{t_2 } {dt_1 \,} e^{ - {{(t_1  - t_I )^2 } \mathord{\left/
 {\vphantom {{(t_1  - t_I )^2 } {2\sigma _1^2 }}} \right.
 \kern-\nulldelimiterspace} {2\sigma _1^2 }}} e^{ - {{(t_2  - t_I )^2 } \mathord{\left/
 {\vphantom {{(t_2  - t_I )^2 } {2\sigma _1^2 }}} \right.
 \kern-\nulldelimiterspace} {2\sigma _1^2 }}} e^{ - i\Omega _I (t_2  - t_I ) + i\Omega _I (t_1  - t_I )}  \hfill \\
  \,\,\,\,\,\,\,\,\,\,\,\,\,\,\,\,\,\,\,\,\,\,\,\,\,\,\,\,\,\,\,\,\,\,\,\,\,\,\,\,\,\,\,\,\,\,\,\, \times \left\langle \left( {\nu _a ,\nu _b } \right)_1 \right| {e^{ - iH_1 (t_I  - t_2 )} e^{ - iH_0 (t_2  - t_I )} e^{ - iH_0 (t_I  - t_1 )} e^{ - iH_1 (t_1  - t_I )} } \left| {\left( {\bar \nu _a ,\bar \nu _b } \right)_1 } \right\rangle  \hfill \\
  \,\,\,\,\,\,\,\,\,\,\,\,\,\,\,\,\,\,\,\,\,\,\,\,\,\,\,\,\,\,\,\,\,\,\,\,\,\,\,\,\,\, = \delta _{\nu _b , \bar \nu _b } \left( {\frac{{iE_I m}}
{2}} \right)^2 \sum\limits_{\bar{\bar \nu} _a } {\left\langle {(\nu _a )_e } \right|} \left. {(\,\bar{\bar \nu} _a )_g } \right\rangle \left\langle {(\,\bar{\bar \nu} _a )_g } \right|\left. {(\,\bar \nu _a )_e } \right\rangle  \hfill \\
  \quad \quad \quad \quad \quad \quad \;\,\,\,\,{\kern 1pt} \,\,\,\,\,\,\,\,\,\,\,\,\, \times I\left( {\Omega _I  - \omega (\bar{\bar \nu} _a  - \bar \nu _a ) - \varepsilon _1 ,\Omega _I  - \omega (\bar{\bar \nu} _a  - \nu _a ) - \varepsilon _1 } \right)\;, \hfill \\
\end{gathered}
\end{equation}
and
	\begin{equation}
\begin{gathered}
  \left\langle \left( {\nu _a ,\nu _b } \right)_{1'} \right| {p_I^{(1'0)} (\infty ;\,t_2 )p_I^{(01)} (t_2 ;\,t_1 )} \left| {\left( {\bar \nu _a ,\bar \nu _b } \right)_1 } \right\rangle  \hfill \\
  \quad \quad \quad \quad \quad \quad \quad  = \left( {\frac{{iE_I m}}
{2}} \right)^2 \int\limits_{ - {\kern 1pt} \infty }^\infty  {dt_2 } \int\limits_{ - {\kern 1pt} \infty }^{t_2 } {dt_1 \,} e^{ - {{(t_1  - t_I )^2 } \mathord{\left/
 {\vphantom {{(t_1  - t_I )^2 } {2\sigma _1^2 }}} \right.
 \kern-\nulldelimiterspace} {2\sigma _1^2 }}} e^{ - {{(t_2  - t_I )^2 } \mathord{\left/
 {\vphantom {{(t_2  - t_I )^2 } {2\sigma _1^2 }}} \right.
 \kern-\nulldelimiterspace} {2\sigma _1^2 }}} e^{ - i\Omega _I (t_2  - t_I ) + i\Omega _I (t_1  - t_I )}  \hfill \\
  \,\,\,\,\,\,\,\,\,\,\,\,\,\,\,\,\,\,\,\,\,\,\,\,\,\,\,\,\,\,\,\,\,\,\,\,\,\,\,\,\,\,\,\,\,\,\,\, \times \left\langle \left( {\nu _a ,\nu _b } \right)_1 \right| {e^{ - iH_{1'} (t_I  - t_2 )} e^{ - iH_0 (t_2  - t_I )} e^{ - iH_0 (t_I  - t_1 )} e^{ - iH_1 (t_1  - t_I )} } \left| {\left( {\bar \nu _a ,\bar \nu _b } \right)_1 } \right\rangle \, \hfill \\
  \,\,\,\,\,\,\,\,\,\,\,\,\,\,\,\,\,\,\,\,\,\,\,\,\,\,\,\,\,\,\,\,\,\,\,\,\,\,\,\,\,\, = \left( {\frac{{iE_I m}}
{2}} \right)^2 \left\langle {(\,\nu _b )_e } \right|\left. {(\,\bar \nu _b )_g } \right\rangle \left\langle {{(\nu _a )_g }}
 \mathrel{\left | {\vphantom {{(\nu _a )_g } {(\,\bar \nu _a )_e }}}
 \right. \kern-\nulldelimiterspace}
 {{(\,\bar \nu _a )_e }} \right\rangle  \hfill \\
  \quad \quad \quad \quad \quad \quad \;\,\,\,\,{\kern 1pt} \,\,\,\,\,\,\,\,\,\,\,\,\, \times I\left( {\Omega _I  - \omega (\nu _a  - \bar \nu _a ) - \varepsilon _1 ,\Omega _I  - \omega (\bar \nu _b  - \nu _b ) - \varepsilon _{1'} } \right)\;. \hfill \\
\end{gathered}
\end{equation}

\
\

\setcounter{section}{3} \setcounter{equation}{0}
\noindent{\large\bfseries \fontfamily{phv}\selectfont
\MakeUppercase{Appendix C: INITIAL ANISOTROPY VALUES FOR PUMP-PROBE AND PUMP-PROBE DIFFERENCE SIGNALS}}

\

For the pump-probe signal, the anisotropy is defined by
	\begin{equation}
r_{{\text{PP}}} (t_{CA} ) = \frac{{{\text{HH - HV}}}}
{{{\text{HH + 2HV}}}}\,\,.
\end{equation}
Using well-known orientational factors$^1$,  we can write the stimulated emission component to the numerator above (assuming the monomers have orthogonal transition dipole moments) as
	\begin{equation}
\begin{aligned}
  {\text{HH - HV}} = &\frac{1}
{{30}}\operatorname{Re} \left( {4\left\langle {{\left\{ {a(10)} \right\}_1 }}
 {\left | {\vphantom {{\left\{ {a(10)} \right\}_1 } {\left\{ {c(10)c(01)a(10)} \right\}_1 }}}
 \right. \kern-\nulldelimiterspace}
 {{\left\{ {c(10)c(01)a(10)} \right\}_1 }} \right\rangle } \right. + 4\left\langle {{\left\{ {a(1'0)} \right\}_{1'} }}
 {\left | {\vphantom {{\left\{ {a(1'0)} \right\}_{1'} } {\left\{ {c(1'0)c(01')a(1'0)} \right\}_{1'} }}}
 \right. \kern-\nulldelimiterspace}
 {{\left\{ {c(1'0)c(01')a(1'0)} \right\}_{1'} }} \right\rangle  \\
  \,\, &- 2\left\langle {{\left\{ {a(1'0)} \right\}_1 }}
 {\left | {\vphantom {{\left\{ {a(1'0)} \right\}_1 } {\left\{ {c(10)c(01)a(1'0)} \right\}_1 }}}
 \right. \kern-\nulldelimiterspace}
 {{\left\{ {c(10)c(01)a(1'0)} \right\}_1 }} \right\rangle  + 3\left\langle {{\left\{ {a(1'0)} \right\}_{1'} }}
 {\left | {\vphantom {{\left\{ {a(1'0)} \right\}_{1'} } {\left\{ {c(1'0)c(01)a(10)} \right\}_{1'} }}}
 \right. \kern-\nulldelimiterspace}
 {{\left\{ {c(1'0)c(01)a(10)} \right\}_{1'} }} \right\rangle  \\
  \,\, &+ 3\left\langle {{\left\{ {a(1'0)} \right\}_1 }}
 {\left | {\vphantom {{\left\{ {a(1'0)} \right\}_1 } {\left\{ {c(10)c(01')a(10)} \right\}_1 }}}
 \right. \kern-\nulldelimiterspace}
 {{\left\{ {c(10)c(01')a(10)} \right\}_1 }} \right\rangle  + 3\left\langle {{\left\{ {a(10)} \right\}_{1'} }}
 {\left | {\vphantom {{\left\{ {a(10)} \right\}_{1'} } {\left\{ {c(1'0)c(01)a(1'0)} \right\}_{1'} }}}
 \right. \kern-\nulldelimiterspace}
 {{\left\{ {c(1'0)c(01)a(1'0)} \right\}_{1'} }} \right\rangle  \\
  \,\, &+ 3\left\langle {{\left\{ {a(10)} \right\}_1 }}
 {\left | {\vphantom {{\left\{ {a(10)} \right\}_1 } {\left\{ {c(10)c(01')a(1'0)} \right\}_1 }}}
 \right. \kern-\nulldelimiterspace}
 {{\left\{ {c(10)c(01')a(1'0)} \right\}_1 }} \right\rangle  - 2\left. {\left\langle {{\left\{ {a(10)} \right\}_{1'} }}
 {\left | {\vphantom {{\left\{ {a(10)} \right\}_{1'} } {\left\{ {c(1'0)c(01')a(10)} \right\}_{1'} }}}
 \right. \kern-\nulldelimiterspace}
 {{\left\{ {c(1'0)c(01')a(10)} \right\}_{1'} }} \right\rangle } \right) \\
   = &\frac{2}
{{30}}\operatorname{Re} \left( {4\left\langle {{\left\{ {a(10)} \right\}_1 }}
 {\left | {\vphantom {{\left\{ {a(10)} \right\}_1 } {\left\{ {c(10)c(01)a(10)} \right\}_1 }}}
 \right. \kern-\nulldelimiterspace}
 {{\left\{ {c(10)c(01)a(10)} \right\}_1 }} \right\rangle } \right. - 2\left\langle {{\left\{ {a(1'0)} \right\}_1 }}
 {\left | {\vphantom {{\left\{ {a(1'0)} \right\}_1 } {\left\{ {c(10)c(01)a(1'0)} \right\}_1 }}}
 \right. \kern-\nulldelimiterspace}
 {{\left\{ {c(10)c(01)a(1'0)} \right\}_1 }} \right\rangle  \\
  \,\, &+ 3\left\langle {{\left\{ {a(1'0)} \right\}_{1'} }}
 {\left | {\vphantom {{\left\{ {a(1'0)} \right\}_{1'} } {\left\{ {c(1'0)c(01)a(10)} \right\}_{1'} }}}
 \right. \kern-\nulldelimiterspace}
 {{\left\{ {c(1'0)c(01)a(10)} \right\}_{1'} }} \right\rangle \, + 3\left. {\left\langle {{\left\{ {a(1'0)} \right\}_1 }}
 {\left | {\vphantom {{\left\{ {a(1'0)} \right\}_1 } {\left\{ {c(10)c(01')a(10)} \right\}_1 }}}
 \right. \kern-\nulldelimiterspace}
 {{\left\{ {c(10)c(01')a(10)} \right\}_1 }} \right\rangle } \right)\,\,\,. \\
  \,\,\,\,\,\, \\
\end{aligned}\end{equation}
In the second equality, use has been made of the fact that for equal-energy homodimers the overlaps retain their value when the labels 1 and $1'$  are interchanged.  The denominator is
	\begin{equation}
\begin{aligned}
  {\text{HH + 2HV}} = &\frac{1}
{3}\operatorname{Re} \left( {\left\langle {{\left\{ {a(10)} \right\}_1 }}
 {\left | {\vphantom {{\left\{ {a(10)} \right\}_1 } {\left\{ {c(10)c(01)a(10)} \right\}_1 }}}
 \right. \kern-\nulldelimiterspace}
 {{\left\{ {c(10)c(01)a(10)} \right\}_1 }} \right\rangle } \right. + \left\langle {{\left\{ {a(1'0)} \right\}_1 }}
 {\left | {\vphantom {{\left\{ {a(1'0)} \right\}_1 } {\left\{ {c(10)c(01)a(1'0)} \right\}_1 }}}
 \right. \kern-\nulldelimiterspace}
 {{\left\{ {c(10)c(01)a(1'0)} \right\}_1 }} \right\rangle  \\
  \,\, &+ \left. {\left\langle {{\left\{ {a(10)} \right\}_{1'} }}
 {\left | {\vphantom {{\left\{ {a(10)} \right\}_{1'} } {\left\{ {c(1'0)c(01')a(10)} \right\}_{1'} }}}
 \right. \kern-\nulldelimiterspace}
 {{\left\{ {c(1'0)c(01')a(10)} \right\}_{1'} }} \right\rangle  + \left\langle {{\left\{ {a(1'0)} \right\}_{1'} }}
 {\left | {\vphantom {{\left\{ {a(1'0)} \right\}_{1'} } {\left\{ {c(1'0)c(01')a(1'0)} \right\}_{1'} }}}
 \right. \kern-\nulldelimiterspace}
 {{\left\{ {c(1'0)c(01')a(1'0)} \right\}_{1'} }} \right\rangle } \right) \\
   = &\frac{2}
{3}\operatorname{Re} \left( {\left\langle {{\left\{ {a(10)} \right\}_1 }}
 {\left | {\vphantom {{\left\{ {a(10)} \right\}_1 } {\left\{ {c(10)c(01)a(10)} \right\}_1 }}}
 \right. \kern-\nulldelimiterspace}
 {{\left\{ {c(10)c(01)a(10)} \right\}_1 }} \right\rangle } \right. + \left. {\left\langle {{\left\{ {a(1'0)} \right\}_1 }}
 {\left | {\vphantom {{\left\{ {a(1'0)} \right\}_1 } {\left\{ {c(10)c(01)a(1'0)} \right\}_1 }}}
 \right. \kern-\nulldelimiterspace}
 {{\left\{ {c(10)c(01)a(1'0)} \right\}_1 }} \right\rangle } \right)\,\,\,. \\
  \,\,\,\,\,\, \\
\end{aligned}
\end{equation}
When the interpulse delay $t_{CA}$ is zero, overlaps like $
\left\langle {{\left\{ {a(1'0)} \right\}_1 }}
 {\left | {\vphantom {{\left\{ {a(1'0)} \right\}_1 } {\left\{ {c(10)c(01)a(1'0)} \right\}_1 }}}
 \right. \kern-\nulldelimiterspace}
 {{\left\{ {c(10)c(01)a(1'0)} \right\}_1 }} \right\rangle
$ vanish, as there is no interval of free evolution during which energy transfer can take place.  This simplification gives a compact expression for the initial anisotropy
	\begin{equation}
r_{{\text{PP}}} (0) = 0.4 + 0.3\frac{{\operatorname{Re} \left\langle {{\left\{ {a(1'0)} \right\}_{1'} }}
 {\left | {\vphantom {{\left\{ {a(1'0)} \right\}_{1'} } {\left\{ {c(1'0)c(01)a(10)} \right\}_{1'} }}}
 \right. \kern-\nulldelimiterspace}
 {{\left\{ {c(1'0)c(01)a(10)} \right\}_{1'} }} \right\rangle }}
{{\operatorname{Re} \left\langle {{\left\{ {a(10)} \right\}_1 }}
 {\left | {\vphantom {{\left\{ {a(10)} \right\}_1 } {\left\{ {c(10)c(01)a(10)} \right\}_1 }}}
 \right. \kern-\nulldelimiterspace}
 {{\left\{ {c(10)c(01)a(10)} \right\}_1 }} \right\rangle }}\, \ \ ,
\end{equation}
which reduces to 0.4 or 0.7 when the ratio of wave-packet overlaps in the second term is 0 or 1, respectively.  This ratio is essentially zero under the conditions of our simulations (which use a red-shifted probe pulse) due to the fact that the spatial region in which the probe pulse is resonant with the $1 \to 0$
 transition (near $q_a  = 2d$
) overlaps the region in which it is resonant with the $1' \leftarrow 0$
 transition ($q_b  = 2d$
) far from the location of the wave packet created by the pump pulse.  Were the pump and probe to have the same center frequency, or were the probe pulse sufficiently short that its center frequency became irrelevant, the value of  $
r_{PP} (0) = 0.7
$
 would be obtained (see Figure panel A).

The anisotropy for the pump-probe difference signal, in which the system first interacts with a vertically polarized control pulse, is given by
	\begin{equation}
r_{{\text{PPD}}} (t_{CA} ) = \frac{{{\text{VHH - VHV}}}}
{{{\text{VHH + 2VHV}}}}\,\,.
\end{equation}
Omitting those overlaps that depend on energy transfer between pulses, and again making use of the symmetry with respect to interchange of 1 and $1'$, we find
\begin{equation}
\begin{gathered}
  {\text{VHH - VHV}} = \frac{2}
{{30}}\operatorname{Re} \left( {2\left\langle {{\left\{ {a(10)} \right\}_1 }}
 {\left | {\vphantom {{\left\{ {a(10)} \right\}_1 } {\left\{ {c(10)c(01)a(10)p(01')p(1'0)} \right\}_1 }}}
 \right. \kern-\nulldelimiterspace}
 {{\left\{ {c(10)c(01)a(10)p(01')p(1'0)} \right\}_1 }} \right\rangle } \right. \hfill \\
  \,\,\,\,\,\,\,\, + \left\langle {{\left\{ {a(1'0)} \right\}_{1'} }}
 {\left | {\vphantom {{\left\{ {a(1'0)} \right\}_{1'} } {\left\{ {c(1'0)c(01)a(10)p(01)p(10)} \right\}_{1'} }}}
 \right. \kern-\nulldelimiterspace}
 {{\left\{ {c(1'0)c(01)a(10)p(01)p(10)} \right\}_{1'} }} \right\rangle  + \left\langle {{\left\{ {a(1'0)} \right\}_{1'} }}
 {\left | {\vphantom {{\left\{ {a(1'0)} \right\}_{1'} } {\left\{ {c(1'0)c(01)a(10)p(01')p(1'0)} \right\}_{1'} }}}
 \right. \kern-\nulldelimiterspace}
 {{\left\{ {c(1'0)c(01)a(10)p(01')p(1'0)} \right\}_{1'} }} \right\rangle  \hfill \\
  \,\,\,\,\,\,\, + 2\left\langle {{\left\{ {a(10)p(01')p(1'0)} \right\}_1 }}
 {\left | {\vphantom {{\left\{ {a(10)p(01')p(1'0)} \right\}_1 } {\left\{ {c(10)c(01)a(10)} \right\}_1 }}}
 \right. \kern-\nulldelimiterspace}
 {{\left\{ {c(10)c(01)a(10)} \right\}_1 }} \right\rangle  + \left\langle {{\left\{ {a(1'0)p(01)p(10)} \right\}_{1'} }}
 {\left | {\vphantom {{\left\{ {a(1'0)p(01)p(10)} \right\}_{1'} } {\left\{ {c(1'0)c(01)a(10)} \right\}_{1'} }}}
 \right. \kern-\nulldelimiterspace}
 {{\left\{ {c(1'0)c(01)a(10)} \right\}_{1'} }} \right\rangle  \hfill \\
  \,\,\,\,\,\,\, + \left. {\left\langle {{\left\{ {a(1'0)p(01')p(1'0)} \right\}_{1'} }}
 {\left | {\vphantom {{\left\{ {a(1'0)p(01')p(1'0)} \right\}_{1'} } {\left\{ {c(1'0)c(01)a(10)} \right\}_{1'} }}}
 \right. \kern-\nulldelimiterspace}
 {{\left\{ {c(1'0)c(01)a(10)} \right\}_{1'} }} \right\rangle } \right) \hfill \\
\end{gathered}
\end{equation}
and
\begin{equation}
\begin{gathered}
  {\text{VHH + 2VHV}} = \frac{2}
{{105}}\operatorname{Re} \left( {9\left\langle {{\left\{ {a(10)} \right\}_1 }}
 {\left | {\vphantom {{\left\{ {a(10)} \right\}_1 } {\left\{ {c(10)c(01)a(10)p(01)p(10)} \right\}_1 }}}
 \right. \kern-\nulldelimiterspace}
 {{\left\{ {c(10)c(01)a(10)p(01)p(10)} \right\}_1 }} \right\rangle } \right. \hfill \\
  \,\,\,\,\,\,\,\, - \left\langle {{\left\{ {a(1'0)} \right\}_{1'} }}
 {\left | {\vphantom {{\left\{ {a(1'0)} \right\}_{1'} } {\left\{ {c(1'0)c(01)a(10)p(01)p(10)} \right\}_{1'} }}}
 \right. \kern-\nulldelimiterspace}
 {{\left\{ {c(1'0)c(01)a(10)p(01)p(10)} \right\}_{1'} }} \right\rangle  - \left\langle {{\left\{ {a(1'0)} \right\}_{1'} }}
 {\left | {\vphantom {{\left\{ {a(1'0)} \right\}_{1'} } {\left\{ {c(1'0)c(01)a(10)p(01')p(1'0)} \right\}_1 }}}
 \right. \kern-\nulldelimiterspace}
 {{\left\{ {c(1'0)c(01)a(10)p(01')p(1'0)} \right\}_1 }} \right\rangle  \hfill \\
  \,\,\,\,\,\,\, + 13\left\langle {{\left\{ {a(10)} \right\}_1 }}
 {\left | {\vphantom {{\left\{ {a(10)} \right\}_1 } {\left\{ {c(10)c(01)a(10)p(01')p(1'0)} \right\}_1 }}}
 \right. \kern-\nulldelimiterspace}
 {{\left\{ {c(10)c(01)a(10)p(01')p(1'0)} \right\}_1 }} \right\rangle  + 9\left\langle {{\left\{ {a(10)p(01)p(10)} \right\}_1 }}
 {\left | {\vphantom {{\left\{ {a(10)p(01)p(10)} \right\}_1 } {\left\{ {c(10)c(01)a(10)} \right\}_1 }}}
 \right. \kern-\nulldelimiterspace}
 {{\left\{ {c(10)c(01)a(10)} \right\}_1 }} \right\rangle  \hfill \\
  \,\,\,\,\,\,\, - \left\langle {{\left\{ {a(1'0)p(01)p(10)} \right\}_{1'} }}
 {\left | {\vphantom {{\left\{ {a(1'0)p(01)p(10)} \right\}_{1'} } {\left\{ {c(1'0)c(01)a(10)} \right\}_{1'} }}}
 \right. \kern-\nulldelimiterspace}
 {{\left\{ {c(1'0)c(01)a(10)} \right\}_{1'} }} \right\rangle  - \left\langle {{\left\{ {a(1'0)p(01')p(1'0)} \right\}_{1'} }}
 {\left | {\vphantom {{\left\{ {a(1'0)p(01')p(1'0)} \right\}_{1'} } {\left\{ {c(1'0)c(01)a(10)} \right\}_1 }}}
 \right. \kern-\nulldelimiterspace}
 {{\left\{ {c(1'0)c(01)a(10)} \right\}_1 }} \right\rangle  \hfill \\
  \,\,\,\,\,\, + 13\left. {\left\langle {{\left\{ {a(10)p(01')p(1'0)} \right\}_1 }}
 {\left | {\vphantom {{\left\{ {a(10)p(01')p(1'0)} \right\}_1 } {\left\{ {c(10)c(01)a(10)} \right\}_1 }}}
 \right. \kern-\nulldelimiterspace}
 {{\left\{ {c(10)c(01)a(10)} \right\}_1 }} \right\rangle } \right)\,\,\,.\,\,\,\,\,\,\,\,\,\, \hfill \\
  \,\,\,\,\,\,\,\,\,\,\,\,\,\,\,\,\, \hfill \\
\end{gathered}
\end{equation}
In the case that all of the overlaps listed in Eqs. (C.6) and (C.7) are equal in value (which happens when the pump and probe are arbitrarily short), the initial anisotropy is 0.7.  Numerically, we find an initial anisotropy of 0.69 when  $\sigma_A$ and  $\sigma_C$ are set equal to one hundredth of the vibrational period for the equal-energy dimer used in Section 2A of the main text (see Figure panels B and C).

When the overlaps containing $c(1'0)c(01)$ do not contribute, for the same reason as described above,  we have
\begin{equation}
\begin{gathered}
  r_{{\text{PPD}}} (0) =  \hfill \\
  \,\frac{7}
{{13 + 9\frac{{\operatorname{Re} \left( {\left\langle {{\left\{ {a(10)} \right\}_1 }}
 {\left | {\vphantom {{\left\{ {a(10)} \right\}_1 } {\left\{ {c(10)c(01)a(10)p(01)p(10)} \right\}_1 }}}
 \right. \kern-\nulldelimiterspace}
 {{\left\{ {c(10)c(01)a(10)p(01)p(10)} \right\}_1 }} \right\rangle  + \left\langle {{\left\{ {a(10)p(01)p(10)} \right\}_1 }}
 {\left | {\vphantom {{\left\{ {a(10)p(01)p(10)} \right\}_1 } {\left\{ {c(10)c(01)a(10)} \right\}_1 }}}
 \right. \kern-\nulldelimiterspace}
 {{\left\{ {c(10)c(01)a(10)} \right\}_1 }} \right\rangle } \right)}}
{{\operatorname{Re} \left( {\left\langle {{\left\{ {a(10)} \right\}_1 }}
 {\left | {\vphantom {{\left\{ {a(10)} \right\}_1 } {\left\{ {c(10)c(01)a(10)p(01')p(1'0)} \right\}_1 }}}
 \right. \kern-\nulldelimiterspace}
 {{\left\{ {c(10)c(01)a(10)p(01')p(1'0)} \right\}_1 }} \right\rangle  + \left\langle {{\left\{ {a(10)p(01')p(1'0)} \right\}_1 }}
 {\left | {\vphantom {{\left\{ {a(10)p(01')p(1'0)} \right\}_1 } {\left\{ {c(10)c(01)a(10)} \right\}_1 }}}
 \right. \kern-\nulldelimiterspace}
 {{\left\{ {c(10)c(01)a(10)} \right\}_1 }} \right\rangle } \right)}}}}\,\,. \hfill \\
\end{gathered}
\end{equation}
In general, this expression gives an initial anisotropy that changes with pulse and molecular parameters, in contrast with the initial value of 0.4 often found in pump-probe experiments.  Were the control pulse arbitrarily short, it would be ineffective in driving nuclear motion in the electronic ground state, and $
\left| {\left\{ {a(10)p(01)p(10)} \right\}_1 } \right\rangle
$
 would be identical to$
\left| {\left\{ {a(10)p(01')p(1'0)} \right\}_1 } \right\rangle
$.  In this limit, Eq. (C.8) reduces to a value of ${7 \mathord{\left/
 {\vphantom {7 {22 \approx 0.32}}} \right.
 \kern-\nulldelimiterspace} {22 \approx 0.32}}$
for non-impulsive pump and probe pulses (see Figure panel B).  That this value for an impulsive control pulse is different from 0.4 is not unexpected.  The control pulse creates a copy of the ground-state nuclear wave packet only in those molecules in which one of the monomers has a nonzero transition dipole component along the V axis.  It is this subset of the total isotropic population upon which a pump-probe experiment is conducted.  In the situation of interest - where the control pulse does generate motion on the electronic ground state - we find no single limiting value for Eq. (C.8) (see Figure panel C).

It should be recalled that our numerical simulations explicitly ignore overlap between the pump and probe pulses, and our results are therefore only strictly valid for interpulse delays longer than the pulse lengths.  Numerical calculations that rigorously include overlap between the pump and probe (not included here) show that neglecting this overlap tends to result in an overestimation of the initial anisotropy for the pump-probe difference.  For example, using the same pulse and molecular parameters as in Fig. 4 of the main body of the paper and neglecting pulse overlap gives an initial anisotropy of 0.25 (which is largely unaffected by inhomogeneous broadening), whereas properly treating the effects of pulse overlap reduces this value to 0.14.  Thus the ``initial anisotropies'' plotted as a function of pump and probe pulse duration do not strictly depict the true initial anisotropy.  Nonetheless, due to the presence of a single vibrational period in our two-mode model system, they do hold significance as the anisotropy when the interpulse delay is equal to the vibrational period, if energy transfer is neglected on this time scale.

\

\noindent \hrulefill \

\noindent $^1$ \footnotesize See, for example, A. Tokmakoff, ``Orientational correlation functions and polarization selectivity for nonlinear spectroscopy of isotropic media. I. Third order,'' J. Chem. Phys. \textbf{105} 1 (1996); A. Tokmakoff, ``Orientational correlation functions and polarization selectivity for nonlinear spectroscopy of isotropic media. II. Fifth order,'' J. Chem. Phys. \textbf{105} 13 (1996). \normalsize
\pagebreak

\begin{tabular}{ m{3.5in} m{0.25in} m{2in} }
\begin{tabular}[t]{m{3.5in}}{\small FIGURE.  Initial anisotropy neglecting pulse overlap (or the anisotropy one vibrational period after the pump pulse arrives, neglecting energy transfer) as a function of pump and probe pulse duration for the equal-energy model system with moderate electron-vibrational coupling ($ \delta ^2  = 2.5$).  The pump pulse is vertically resonant at the ground-state equilibrium geometry, while the probe is resonant at the outer turning point for nuclear motion in the excited electronic state ($\Omega _A  = \varepsilon _1  + \delta ^2 \omega$ and $\Omega _C  = \varepsilon _1  - 3\delta ^2 \omega$).  A) The pump-probe anisotropy takes a value of 0.7 for an impulsive probe, and 0.4 otherwise.  B) The pump-probe difference anisotropy using an impulsive control pulse, showing the same behavior as (A) but with a lower limit of 7/22 rather than 0.4.  C) The pump-probe difference anisotropy using a control pulse optimized to generate coherent ground-state nuclear motion (as described in the main text) shows more complicated behavior than (A) or (B).  The anisotropy is high for an impulsive probe, however as the probe pulse becomes longer the initial anisotropy does not decrease monotonically. \normalsize } \end{tabular} & & \begin{tabular}{c} \includegraphics[scale=0.4]{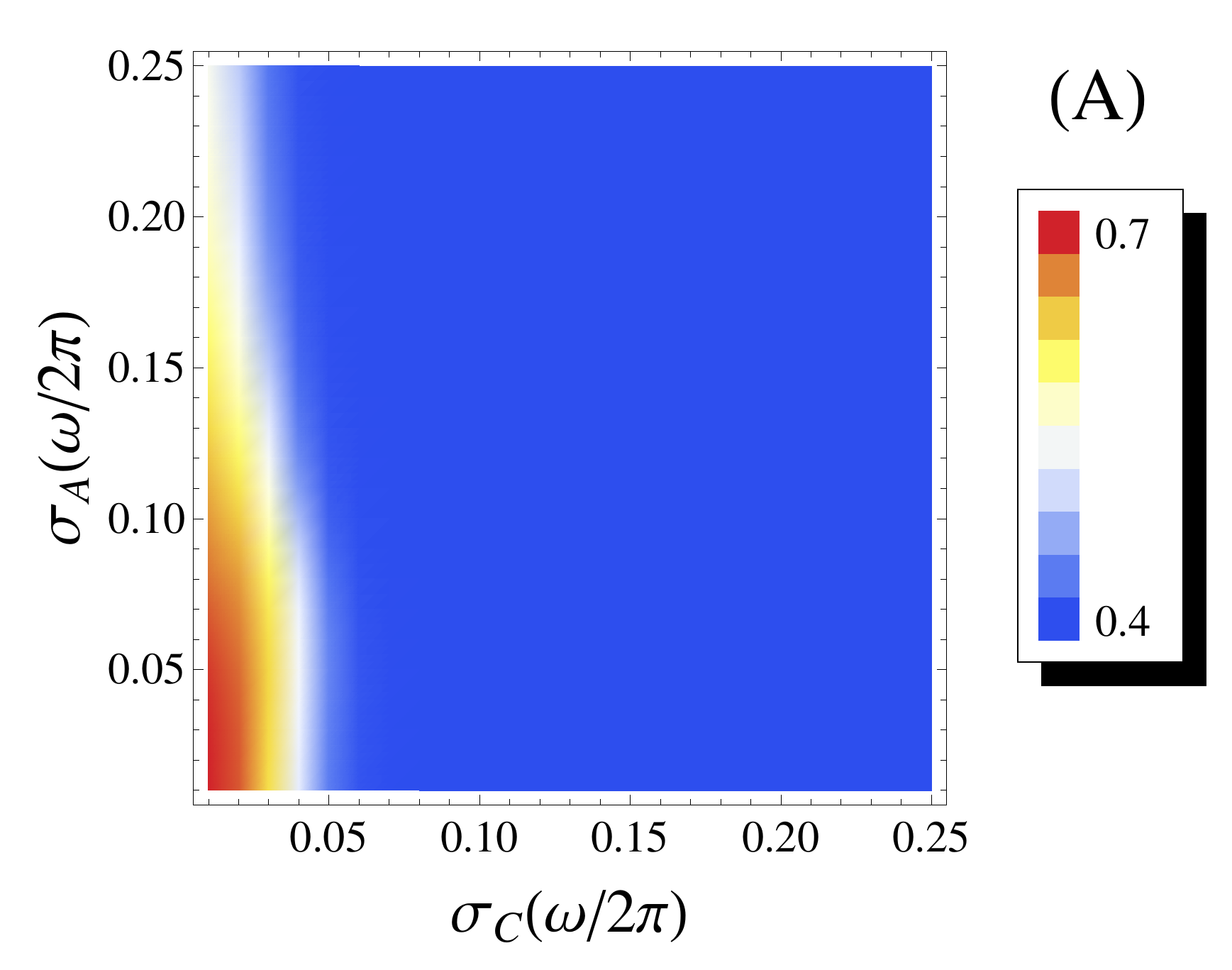} \\ \includegraphics[scale=0.4]{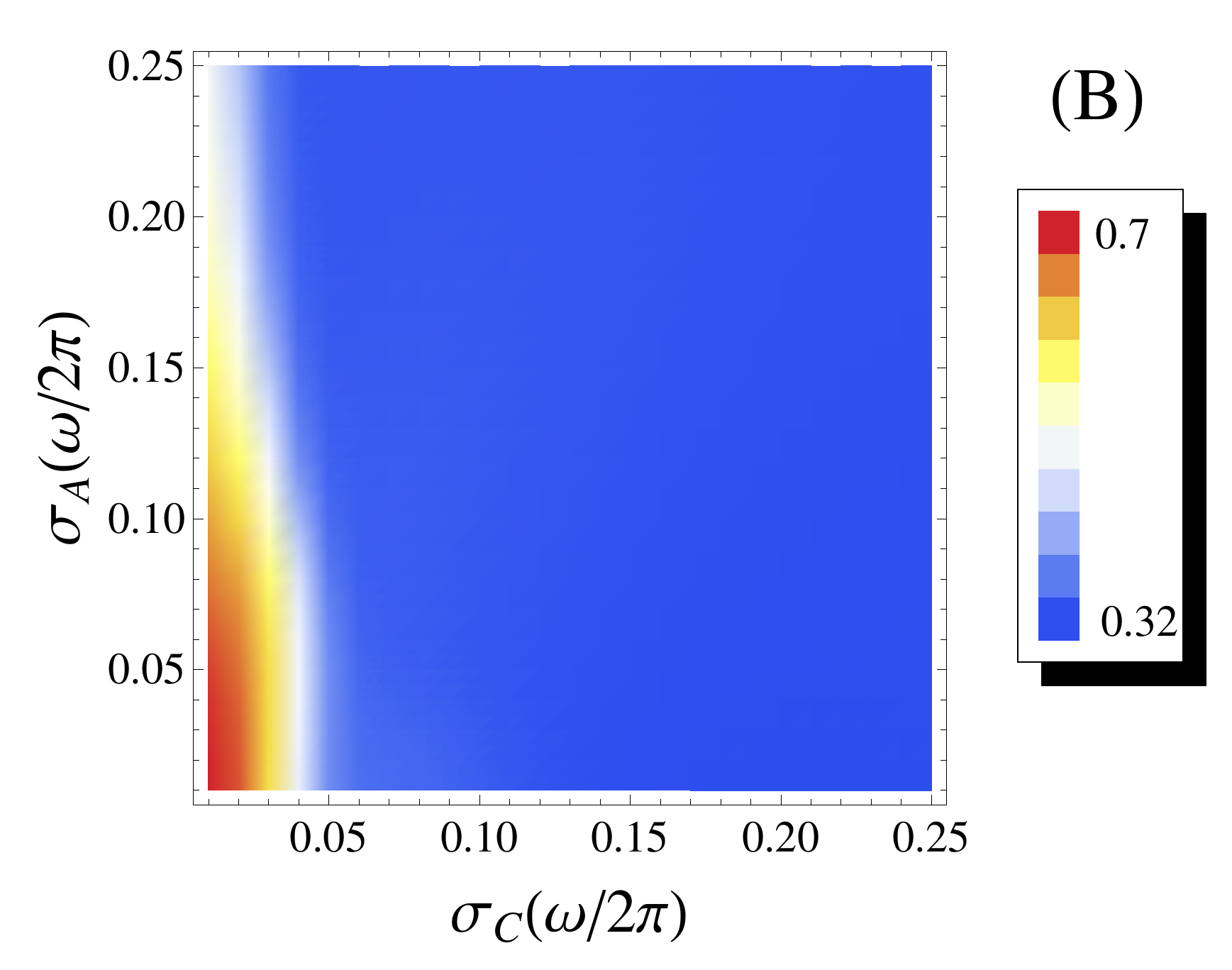} \\ \includegraphics[scale=0.4]{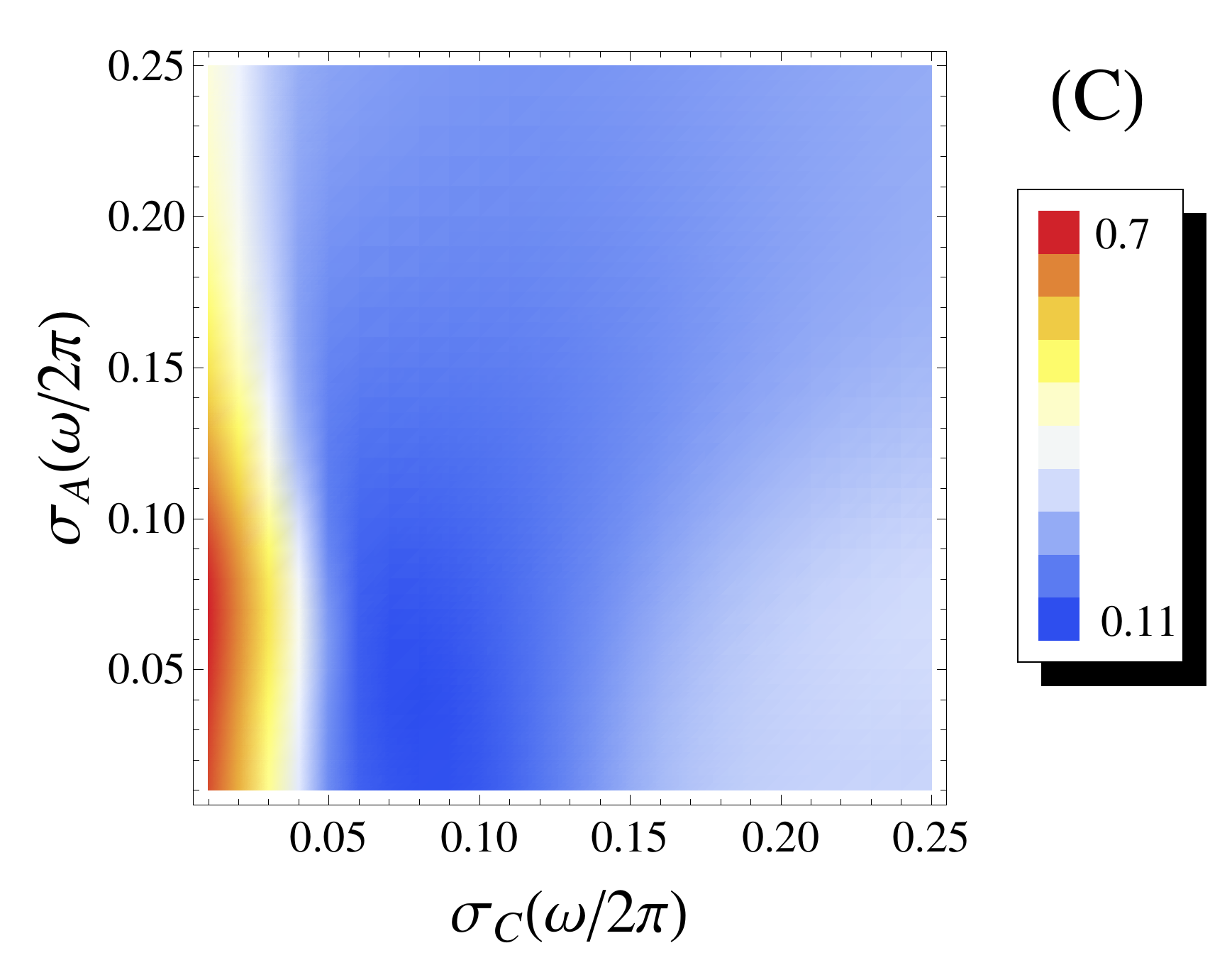} \end{tabular}
\end{tabular}

\end{document}